\documentclass{emulateapj}
\usepackage{apjfonts}
\usepackage{graphicx}
\usepackage[space]{grffile}
\usepackage{latexsym}
\usepackage{amsfonts,amsmath,amssymb}
\usepackage{url}
\usepackage[utf8]{inputenc}
\usepackage{fancyref}
\usepackage{hyperref}
\hypersetup{colorlinks=True,pdfborder={0 0 0},breaklinks,colorlinks,citecolor=blue,linkcolor=magenta}
\usepackage{textcomp}
\usepackage{longtable}
\usepackage{multirow,booktabs}
\usepackage{float}

\newcommand{\Rv}{R$_{vir}$}
\newcommand{\gal}{{\tt GALFIT}}
\newcommand{\sx}{{\tt SExtractor}}
\newcommand{\iraf}{{\tt IRAF}}

\newcommand{\z}{$\bar{z}=0.92$}
\newcommand{{\HST}}{\emph{HST}}
\newcommand{{\M}}{log($M_{*}$/M$_{\odot}$)}
\newcommand{{\kpc}}{~kpc}
\newcommand{{\n}}{S\'{e}rsic}
\newcommand{{\s}}{M$_{\odot}$~yr$^{-1}$}
\newcommand{\rad}{$2\textgreater$R$_{vir}\textgreater0.5$}
\newcommand{\radin}{R$_{vir}$$\textless0.5$}
\newcommand{\ms}{mass$-$size}
\newcommand{\y}{yr$^{-1}$}

\shorttitle{The effects of environment at $z\sim1$}
\shortauthors{R. Allen et al.}

\begin{document}

\title{Differences in the structural properties and star-formation rates of field and cluster galaxies at $z\sim1$}

\author{Rebecca J. Allen\altaffilmark{1,3},
Glenn G. Kacprzak\altaffilmark{1},
Karl Glazebrook\altaffilmark{1},
Kim-Vy H. Tran\altaffilmark{2},
Lee R. Spitler\altaffilmark{3,4},
Caroline M. S. Straatman\altaffilmark{5},
Michael Cowley\altaffilmark{3,4},
Themiya Nanayakkara\altaffilmark{1}
}
                                                                                
\altaffiltext{1}{Swinburne University of Technology, Victoria 3122,
Australia}
\altaffiltext{2}{George P. and Cynthia Woods Mitchell Institute for Fundamental Physics and Astronomy, and Department of Physics and Astronomy, Texas A\&M University, College Station, TX 77843}
\altaffiltext{3}{Australian Astronomical Observatories, PO Box 915, North Ryde, NSW 1670, Australia}
\altaffiltext{4}{Department of Physics \& Astronomy, Macquarie University, Sydney, NSW 2109, Australia}
\altaffiltext{5}{Leiden Observatory, Leiden University, P.O. Box 9513, 2300 RA Leiden, The Netherlands}

\begin{abstract}
We investigate the dependance of galaxy sizes and star-formation rates (SFRs) on environment using a mass-limited sample of quiescent and star-forming galaxies with {\M}$\geq9.5$ at {\z} selected from the NMBS survey. 
Using the GEEC2 spectroscopic cluster catalog and the accurate photometric redshifts from NMBS, we select quiescent and star-forming cluster ($\bar{\sigma}=490$~km~s$^{-1}$) galaxies within two virial radius, {\Rv}, intervals of {\rad} and {\radin}. 
Galaxies residing outside of 2~{\Rv} of both the cluster centres and additional candidate over-densities are defined as our field sample.
Galaxy structural parameters are measured from the COSMOS legacy {\HST}/ACS F814W image.  
The sizes and {\n} indices of quiescent field and cluster galaxies have the same distribution regardless of {\Rv}. 
However, cluster star-forming galaxies within 0.5~{\Rv} have lower mass$-$normalised average sizes, by $16\pm{7}\%$, and a higher fraction of {\n} indices with $n\textgreater1$, than field star-forming galaxies.  
The average SFRs of star-forming cluster galaxies show a trend of decreasing SFR with clustocentric radius.
The mass$-$normalised average SFR of cluster star-forming galaxies is a factor of $2-2.5$ ($7-9\sigma$) lower than that of star-forming galaxies in the field.
While we find no significant dependence on environment for quiescent galaxies, the properties of star-forming galaxies are affected, which could be the result of environment acting on their gas content.
\end{abstract}

\section{Introduction}
The size growth rate of galaxies is indicative of the mechanisms that drive galaxy evolution.
Quiescent galaxies with {\M}$\sim11$ have demonstrated an accelerated growth having sizes $4-6$ times larger at $z=0$ compared to $z=2-4$, while star-forming galaxies with similar masses have only grown by a factor of 2 since $z=4$ \citep{2014ApJ...785...18M,2014ApJ...788...28V,2015ApJ...808L..29S}.

The dramatic size growth of quiescent galaxies is partly attributed to their decedent nature; some fraction of their population form when already massive star-formers collide in major mergers or quench.
Once they have formed, it is thought that the majority of their size growth comes from adiabatic expansion \citep[e.g.,][]{2008ApJ...689L.101F,2010ApJ...718.1460F} or minor and major merging events \citep[e.g.,][]{2006ApJ...648L..21K,2009ApJ...699L.178N,Guo2011,2013MNRAS.429.2924H,2013MNRAS.428.1088M,2013ApJ...763...73S}.
However, studies have shown that the rate of minor mergers at $z\textgreater1$ is not sufficient to be the dominant means of growth for passive galaxies \citep{2012ApJ...746..162N,2014ApJ...783..117B}.
Therefore, the mechanisms that drive the accelerated growth of quiescent galaxies are still not well understood.

On the other hand, the steady growth of star-forming galaxies is attributed mainly to the production of new stars from cold gas reservoirs and inflowing streams \citep[e.g.,][]{2012A&A...544A..68L,2013Sci...341...50B}, and possibly via minor mergers \citep{vanDokkum2010}.
While observations and simulations have provided some insight into the channels that drive galaxy growth, one important aspect is still debated: the role of environment.

At $z=0$, there is a clear relation between the morphology of galaxies and the environment they inhabit \citep{1980ApJ...236..351D,1997ApJ...490..577D} with elliptical galaxies being more prevalent in high density regions. 
It has become apparent that as early as $z=2-3$ these large scale structures began to form \citep{2014A&A...572A..41L,2012ApJ...745..106L,2012ApJ...748L..21S}.
Therefore, the role of environment must be understood to better constrain the instruments that drive galaxy growth.

It is thought that high density environments are efficient at removing the cold gas from star-forming galaxies via galaxy-galaxy interactions, strangulation, and harassment \citep{1972ApJ...176....1G,2007ASPC..379..243M}.
The depletion of cold gas and quenching of star-forming galaxies, could stunt their size growth, creating an observable difference between their sizes/star-formation rates and those of field star-forming galaxies.
In fact, the star-formation density relation has been observed from $z\sim0-2$ \citep[e.g.,][]{2004MNRAS.353..713K,2011ApJ...735...53P,2012ApJ...744...88Q}, providing direct evidence that environment is effective at quenching star-forming galaxies.

However, the effects of environment on star-forming galaxies may not be completely destructive.
At $z\geq1$ the cores of clusters are hosts to galaxies with star-formation rates (SFRs) up to $\sim100~$M$_{\odot}$ yr$^{-1}$ \citep{2008MNRAS.383.1058C,2010ApJ...718..133H,2010ApJ...719L.126T,2012ApJ...745..106L}.
The elevated SFRs of these galaxies could lead to bulge growth and a transformation from late-type morphologies to bulge dominated early-type morphologies.
In fact, \citet{2015ApJ...804..117M} have confirmed a substantial population of star-forming early-type galaxies clusters at $z=1.84$ and $z=1.9$.
The {\ms} relation of these cluster galaxies follows that of passive early-type cluster galaxies at $z\sim0.7-1.5$.
Furthermore, \citet{2014ApJ...788...11L} found that star-forming galaxies with {\M}$\textgreater11$ have bulge to total ratios (B/T) of $40-50\%$.
Assuming imminent quenching, then significant bulge growth and a transformation of late-type to early-type is expected for massive star-formers transitioning into quiescence. 
Therefore, measuring the properties of star-forming galaxies at different cluster radii may provide answers to the level of impact environment plays on their size growth.

Simulations \citep[e.g.,][]{2014MNRAS.439.3189S} indicate a strong dependence of median galaxy size on halo mass with quiescent galaxies in higher mass halos having larger sizes by a factor of $\sim1.5-3$ .
If quiescent galaxies are undergoing accelerated growth due to higher merger rates in clusters, then a measurable size difference should be present.
Therefore, it is important to compare the sizes of quiescent galaxies in different environments to understand if their accelerated growth can be attributed to major merger events. 

To date, there are a number of studies that have used a combination of high resolution imaging, multi-band photometry, and spectroscopy to study the stellar mass-size relation as a function of environment.
These studies span $0\leq{z}\leq2$ and show that the effect of environment on the sizes of quiescent galaxies is either weak or non-existent \citep{Papovich2012,2013ApJ...770...58B,2013ApJ...779...29H,2013MNRAS.428.1715H,2014MNRAS.444..682C,2014ApJ...788...51N,2015ApJ...806....3A,2015MNRAS.450.1246K}.

There are few studies that have examined the effect of environment on star-forming galaxy sizes up to $z\sim2$.
At $z=0.12$, \citet{2014MNRAS.444..682C} find that field late-type galaxies with {\M}$\sim10.3$ are up to $7.5\%$ larger in size than cluster late-type galaxies of similar mass.
At $0.4\textless{z}\textless0.8$, \citet{2015MNRAS.450.1246K} find no significant difference in the sizes of field and cluster late-type galaxies with {\M}$\textgreater10.2$.
At $z=2.1$, \citet{2015ApJ...806....3A} found that the mass$-$normalised sizes of star-forming cluster galaxies with {\M}$\geq9$, are $12\%$ larger than their field counter-parts.
The conflicting results and lack of a strong size difference could be due to an evolution in the SFR-density relation as well as differences in galaxy sample selection.

To understand if there is truly a size dependence on environment, it is important to determine at what epoch these differences emerged and to trace their evolution.
However, finding and quantifying environment at $z\geq0.8$ is difficult because accurate redshifts are necessary, and spectroscopy becomes time expensive.

The Galaxy Environment Evolution Collaboration 2 (GEEC2) spectroscopic survey \citep{2014MNRAS.443.2679B} has identified 11 galaxy clusters in the COSMOS field with $0.8\textless{z}\textless1$ ($\bar{z}=0.82$, $\bar{\sigma}=380$~km~s$^{-1}$).
These clusters were found as part of a follow up survey that utilises the spectroscopic catalog of zCOSMOS \citet{2007ApJS..172...70L} and the X-ray catalog of \citet{2011ApJ...742..125G}.
With the use of spectroscopy, \citet{2014MNRAS.443.2679B} have confirmed over-densities that can be used to probe the effects of environment on galaxy evolution.

While spectroscopy produces accurate redshifts, it is biased towards bright and/or blue objects.
The photometric data obtained from ground based surveys such as the NEWFIRM medium-band Survey (NMBS), can be used to calculate very accurate photometric redshifts, rest-frame colors, stellar masses, SFRs, and other galaxies properties \citep[e.g.,][]{vanDokkum2010,2011ApJ...743..168K,2012ApJ...754L..29W,Whitaker2012}. 
Therefore, it can be used to create more complete samples of both star-forming and quiescent galaxies.

Lastly, it is vital to have high resolution imaging from which to measure the physical properties of galaxies, such as size.
Through legacy surveys, such as CANDELS \citep{2011ApJS..197...35G,2011ApJS..197...36K}, public {\HST} imaging provides coverage in multiple wavelengths of several legacy fields.
The PSF FWHM of {\HST} imaging ranges from $0.16-0.10''$, therefore, it is possible to measure galaxy sizes down to $\sim0.08-0.05''$ (or $\sim0.5$ kpc at $z=1$) \citep{vanDokkum2010}.

In this paper, we use both the GEEC2 survey and NMBS to select a mass-complete ({\M}$\geq9.5$) sample of field and cluster galaxies.
For the first time, we compare the structural properties of galaxies as a function of cluster virial radius to quantify where the effects of environment begin.
We use the virial radius to determine field, cluster outskirt, and cluster core samples.
To understand if environment is affecting the growth of star-forming and quiescent galaxies, we separate galaxies based on their star-formation activity using rest-frame colors.
We use {\HST}/ACS F814W imaging to measure the structural parameters of our samples of galaxies to compare the {\ms} relation and average {\n} indices of field and cluster, star-forming and quiescent galaxies at $z\sim1$.
To complement the {\ms} relation, we analyse the average SFRs of our sample to gain additional insight on how galaxy growth may be affected by environment.

The paper is organised as follows: in Section \ref{sec:sam} we describe our sample selection and its properties, in Section \ref{sec:ana} we describe our construction of the mass-size relation, followed by our results regarding the average sizes, {\n} indices, and SFRs in Section \ref{sec:res}.
We discuss the consequences of our findings in Sections \ref{sec:dis} and \ref{sec:con}.
We assume a $\Lambda$CDM cosmology with $\Omega_{\Lambda}=0.73$, $\Omega_{m}=0.27$, and $H_{0}=71$ km s$^{-1}$.
\section{Sample}
\label{sec:sam}
\subsection{Cluster Locations}
We use the GEEC2 spectroscopic survey \citep{2014MNRAS.443.2679B} to obtain galaxy clusters at $z\sim1$.
In the GEEC2 survey, candidate clusters were chosen from the X-ray catalog of \citet{2011ApJ...742..125G} and from the previous spectroscopic survey zCOSMOS \citep{2007ApJS..172...70L}.
To confirm the existence of candidate clusters, \citet{2014MNRAS.443.2679B} searched for additional galaxy members by obtaining Gemini/GMOS spectra of galaxies with photometric redshifts from \citet{2007ApJS..172...99C} that were consistent with the clusters.

The GEEC2 cluster centres were chosen based on the original X-ray centres of \citet{2011ApJ...742..125G} and confirmed by the mean location of all spectroscopic members of the cluster.
Velocity dispersions were calculated using the spectroscopic members and range between 350 and 690~km~s$^{-1}$.
From the velocity dispersions they calculated the virial radii, {\Rv}.

We use the cluster centres and virial radii defined by the spectroscopically confirmed galaxies to select our photometric sample of cluster galaxies, which is outlined in Section \ref{sec:fc}.
For more details regarding cluster centres, velocity dispersions, and their observations, please refer to \citet{2014MNRAS.443.2679B}.
\subsection{Photometric Catalog}
\label{sec:phots}
For our sample of field and cluster galaxies we use NMBS \citep{2011ApJ...735...86W} to obtain accurate photometric redshifts, stellar masses, rest-frame colors, and star-formation rates (SFRs).
As outlined above, NMBS is highly advantageous because it is a deep survey that utilises medium-band photometry that can be well fit by template SEDs, providing very accurate redshifts and galaxy properties, such as mass, SFR, stellar ages, etc., without the use of spectroscopy.

NMBS stellar masses were obtained by fitting stellar population templates to the photometric data using the code {\tt FAST} \citep{Kriek2009}.
Stellar population models were made with the population synthesis code of \citet{BruzualCharlot2003} assuming a Chabrier IMF and solar metallicity.
Star formation histories were modelled as exponentially decreasing ($\Psi~{\propto}~e^{-t/\tau}$) with values of log($\tau$/year)$=7-10$ in steps of 0.2 and log(age/yr)$=7.6-10.1$ in steps of 0.1. 
NMBS is mass complete down to {\M}$=9.5$ at $z=1$.
SFRs were derived using UV and IR luminosities, which includes MIPS 24 micron photometry and the rest-frame 2800{\AA} luminosity, and is outlined in \citet{2012ApJ...754L..29W}.
NMBS redshifts and rest-frame colors were determined by fitting template SEDs with the code {\tt EAZY} \citep{Brammer2008}.
The accuracy of the photometric redshifts of NMBS is $\sigma_{z}$/($1+z$)$\textless0.015$ and we will refer to it as $\sigma$ for the rest of the paper.
For a full outline of the parameters and models that were used in both {\tt EAZY} and {\tt FAST}, please see \citet{2011ApJ...735...86W}.

NMBS has one square degree pointing in the COSMOS and AEGIS fields.
For this study we use data that covers COSMOS to match galaxies from GEEC2.

\subsection{Field and Cluster Sample Selection}
\subsubsection{Cluster Galaxies}
Due to the biases of spectroscopy towards star-forming galaxies, we select galaxies using the accurate photometric redshifts of NMBS ($\sigma_{z}$/($1+z$)$=0.015$) instead of using only spectroscopic cluster members from GEEC2.
The NMBS footprint differs from that of the GEEC2 survey, therefore, we use three clusters (GEEC2 IDs:130,143, and 150) that are situated in both NMBS and GEEC2 footprints. 
The total number of GEEC2 cluster galaxies in NMBS is 93.
The spectroscopic redshift range of the three clusters is $0.83-0.94$ ({\z}).

We select cluster galaxies from the COSMOS \citep{2007ASPC..375..166S} field, within $\pm~4\sigma$ of the spectroscopic redshift of each cluster centre.
The 4$\sigma$ selection guarantees that we recover $94\%$ (87/93) of the original GEEC2 spectroscopic cluster members in NMBS using photometric redshifts. 
\begin{figure}[h]
\epsscale{1.2}\plotone{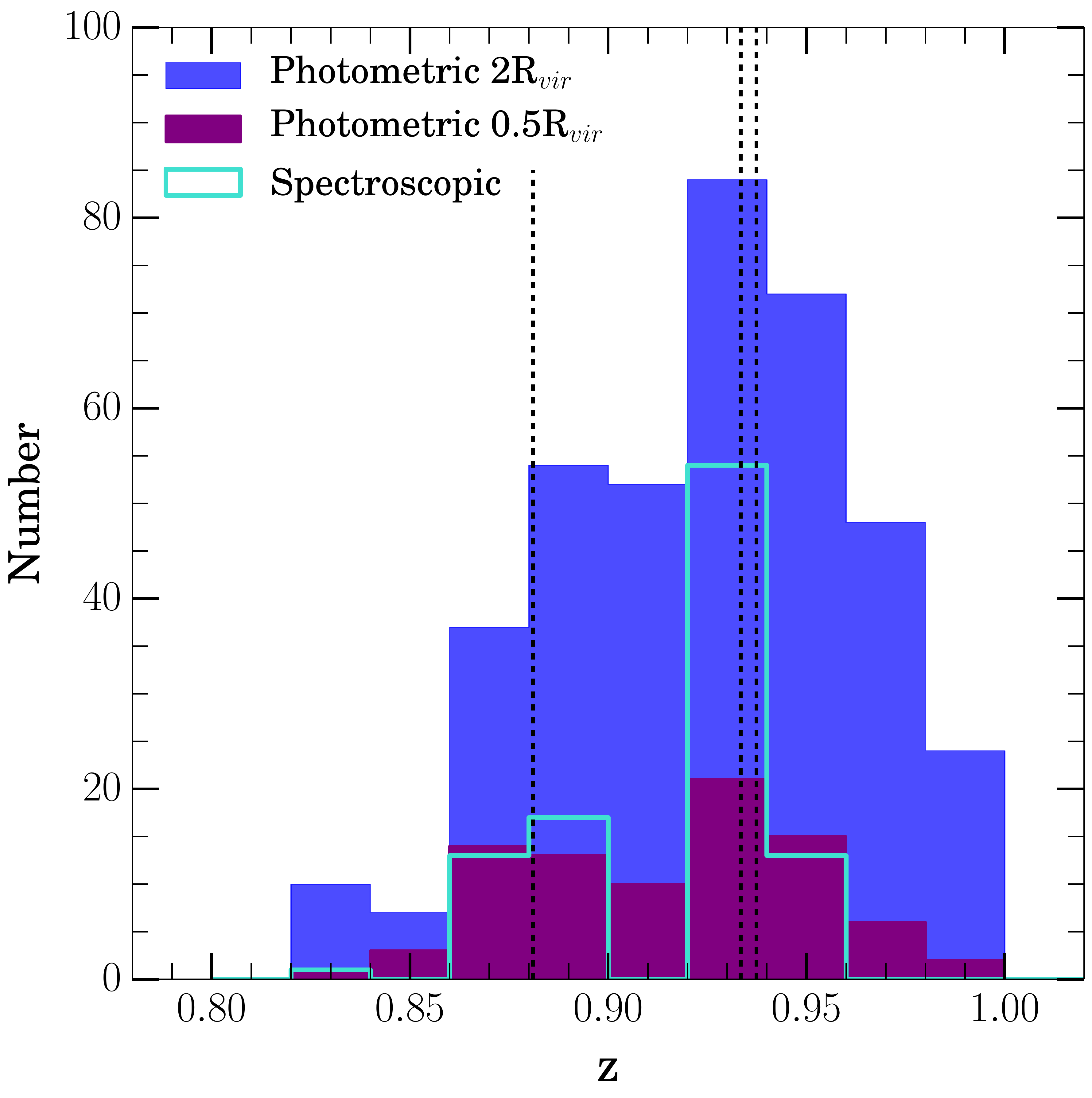}
  \caption{The distribution of redshifts for our photometrically selected cluster samples compared to the spectroscopic redshifts of the GEEC2 sample that resides in NMBS. The cluster spectroscopic redshifts are denoted by black dashed lines.}
  \label{fig:zs}
\end{figure}
\label{sec:fc}
\begin{figure*}[]
\epsscale{1.2}\plotone{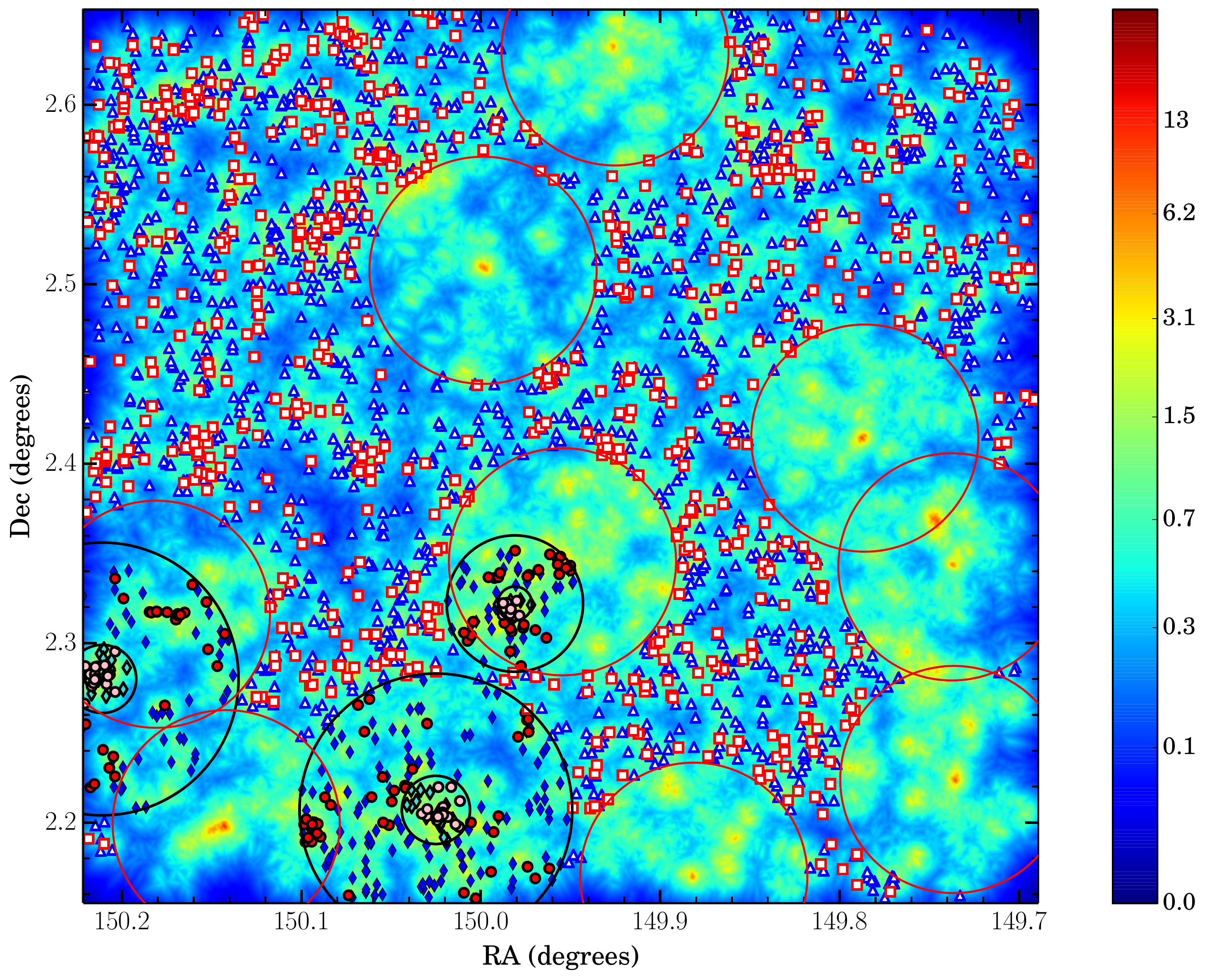}
  \caption{Seventh-nearest-neighbor projected density map of the NMBS footprint in the COSMOS field. The color bar represents the significance, in sigma, of the projected density at $0.76\textless{z}\textless1.06$ above the mean density. The mean density is averaged over the field at $0.46\leq{z}\leq0.76$ and $1.06\leq{z}\leq1.36$. Field star-forming galaxies (open blue triangles) and field quiescent galaxies (open red squares) were selected where no significant ($\textgreater15\sigma$) large-scale candidate over-densities were found. These additional over-densities are denoted by red circles, and have radii $=227''$. Three galaxy over-densities in the COSMOS field were found in the GEEC2 survey \citep{2014MNRAS.443.2679B}, and confirmed using the seventh-nearest-neighbor metric, with photometric redshifts between $0.821\textless{z}\textless1.004$. The cluster samples are selected within 2~{\Rv} and 0.5~{\Rv}, $200-100''$($\sim~1.05-0.53$ Mpc at {\z}) and $67-33''$($\sim~0.53-0.26$ Mpc at {\z}), respectively, of each cluster centre (black solid circles). Our samples of 2~{\Rv} cluster star-forming and quiescent galaxies are shown as blue diamonds and red circles, respectively. The 0.5~{\Rv} cluster samples have the same symbols but are a lighter color.}
  \label{fig:dens}
\end{figure*}
We select two cluster samples using {\Rv}.
We use {\rad}, which we refer to as 2~{\Rv} for the rest of the paper, and {\radin}, which we refer to as 0.5~{\Rv} for the rest of the paper.
For massive galaxy haloes, $M_h\geq10^{14}$ M$_{\odot}$, substantial populations of red galaxies have been observed out to 2-4{\Rv} \citep{2009ApJ...699.1333H}.
Because the cluster haloes of GEEC2 are on the order of $M_h\sim10^{13-14}$ M$_{\odot}$, we create an outer (2~{\Rv}) cluster sample to understand if environment is affecting the sizes of galaxies at larger radii. 
The second radius is used to create a core cluster sample that is within 0.5~{\Rv} of each cluster centre.
0.5~{\Rv} is generally considered to be the transition radius between the cluster core and outskirts \citep[e.g.,][]{2014MNRAS.441..203D}.

The physical sizes of the apertures for the two different selections range $200-100''$($\sim~1.05-0.53$ Mpc at {\z}) and $67-33''$($\sim~0.53-0.26$ Mpc at {\z}), for 2~{\Rv} and 0.5~{\Rv}, respectively.
The total fraction of GEEC2 spectroscopically confirmed galaxies that reside in NMBS to our photometrically selected galaxies within these two radii are 42/50 and 37/37 for 2~{\Rv} and 0.5~{\Rv}, respectively.
The distribution of redshifts for our photometrically selected samples compared to the GEEC2 spectroscopic sample are shown in Figure~\ref{fig:zs}.

We use masses from NMBS to create a mass complete sample by including all galaxies with {\M}$\textgreater9.5$. 
The total number of galaxies in each of our samples is listed in Table~\ref{table:sam}.

\subsubsection{Field Galaxies}
The field sample was selected from the COSMOS field using the NMBS photometric redshifts. 
To locate any other significant over-densities besides the GEEC2 clusters, we use the seventh-nearest-neighbor technique to map the projected density of the COSMOS field within $0.76\textless{z}\textless1.06$. 
This redshift range corresponds to $\pm~8\sigma$ of the lowest and highest spectroscopic cluster centres. 
The significance of any over densities is determined by comparing the mean density at $0.46\leq{z}\leq0.76$ and $1.06\leq{z}\leq1.36$ to the mean density at $0.76\textless{z}\textless1.06$. 

In Figure~\ref{fig:dens}, we show the seventh-nearest-neighbour projected density map of the COSMOS field at $0.76\textless{z}\textless1.06$ overlaid with the field and cluster samples.
The star-forming 2 and 0.5~{\Rv} cluster galaxies, detected for our sample, are shown as solid blue and turquoise diamonds, respectively.
The quiescent 2 and 0.5~{\Rv} cluster galaxies, detected for our sample, are shown as solid red and pink circles, respectively.
The quiescent field and star-forming galaxies are shown as red open squares and blue open triangles, respectively.
We discuss the separation of star-forming and quiescent galaxies in Section~\ref{sec:uvj}.

Our seventh-nearest-neighbour technique recovers all GEEC2 clusters, and their significance is $\geq15\sigma$. 
We note that additional over-densities are present in the NMBS field.
However, we can neither confirm that these candidate over-densities are actual clusters instead of chance projects, or can adequately characterized them (i.e., determine their {\Rv}, velocity dispersions, or halo masses), so we do not include them in our analysis.
These extra over-densities, of significance greater than $15\sigma$, are shown as red circles in Figure~\ref{fig:dens}. 
The size of the red circles are taken as the average 2~{\Rv} of our GEEC2 clusters, $\sim230''$.
Our field samples are selected outside of these apertures to ensure that they do not reside in over-dense regions.

We use the redshift bin, $0.8\leq{z}\leq1$, to select our field galaxies, which corresponds to $\pm~4\sigma$ of the lowest and highest spectroscopic redshift of the cluster centres.
We use this narrower redshift bin to eliminate any overlap with neighbouring over-densities.
We use the same mass limit as for the cluster samples including all galaxies with {\M}$\textgreater9.5$.
The total number of galaxies in each of our samples is listed in Table~\ref{table:sam}.
\begin{table}[]
\begin{center}
\caption{Sample Sizes}
\label{table:sam}
	\begin{tabular}{c c c c}
	\hline\hline
	& Field & {\rad} Cluster & {\radin} Cluster  \\
	\hline\\[-1.5ex]
	Star-forming & 1189 & 206 & 38 \\
	\hline\\[-1.5ex]
	Quiescent & 535 & 97 & 47\\
	\hline \\
	\end{tabular}
	\end{center}
\end{table}
\subsection{Separating Quiescent and Star-Forming Galaxies}
\label{sec:uvj}
\begin{figure}[]
\epsscale{1.2}\plotone{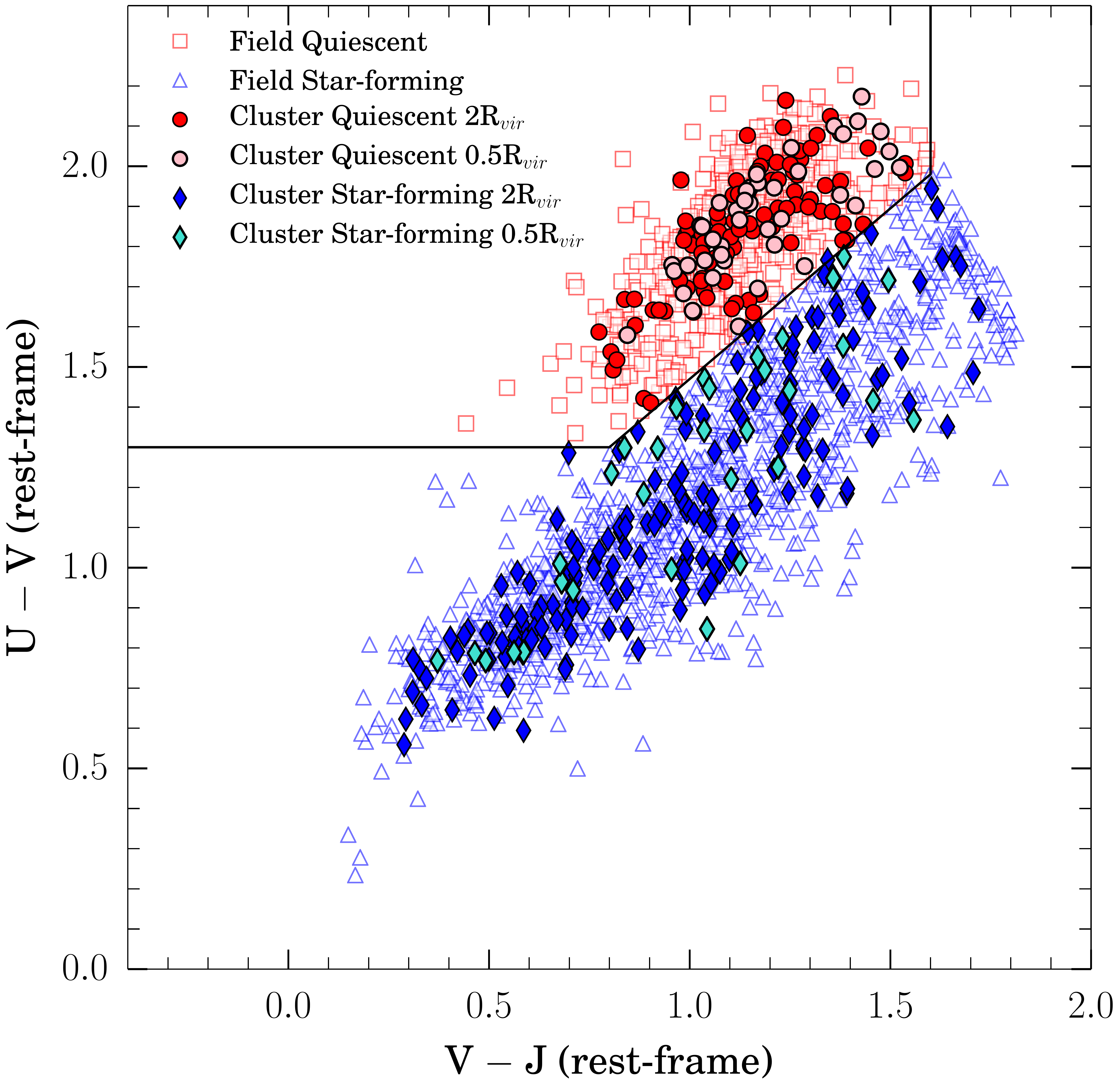}
  \caption{Rest-frame U~$-$~V versus V~$-$~J colors for our sample of field and cluster galaxies at {\z}. Star-forming cluster (field) galaxies are shown as filled (open) blue diamonds (triangles). Quiescent 2~{\Rv} cluster (field) galaxies are shown as filled (open) red circles (squares). The cluster galaxies chosen within 0.5~{\Rv} are the same symbol as their 2~{\Rv} counter-parts but a lighter color. The black line represents the boundary for quiescent galaxies (above) and star-forming galaxies (below). We use this diagram to separate our sample into star-forming and quiescent galaxies.}
  \label{fig:uvj}
\end{figure}
\begin{figure}[]
\epsscale{1.2}\plotone{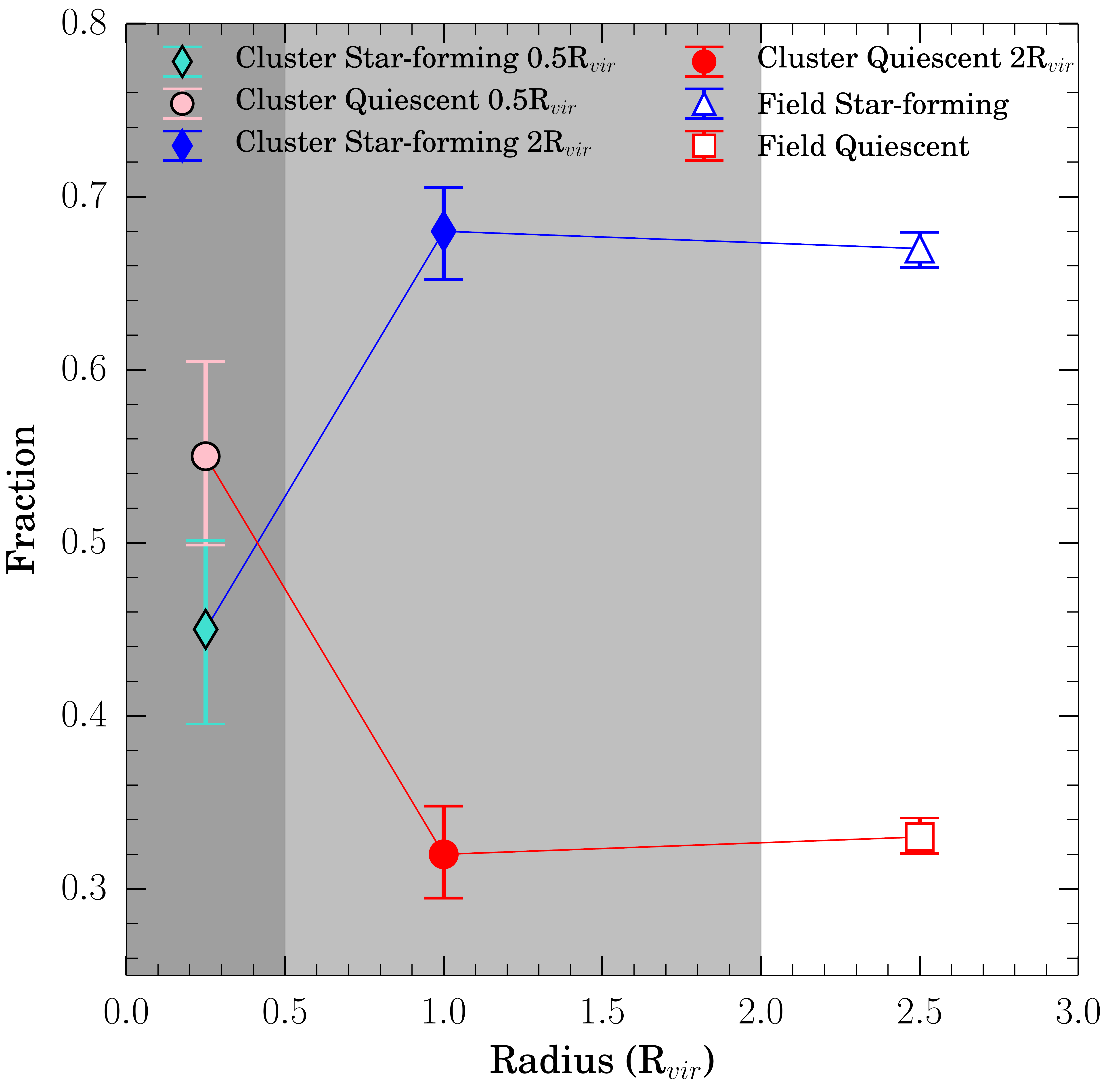}
  \caption{The fraction of quiescent (red and pink points) and star-forming galaxies (blue and turquoise points) as a function of cluster radius. The grey contours represent the different radii that samples are chosen from. The fraction of quiescent galaxies increases from $33\%$ to $55\%$ from the field to the cluster core.}
  \label{fig:rfrac}
\end{figure}
Due to the known difference in size evolution between star-forming and quiescent galaxies \citep[e.g.,][]{2014ApJ...788...28V}, it is important to separate these two populations before we can determine their {\ms} relations.
We use the U~$-$~V versus V~$-$~J rest-frame color-color diagram to separate quiescent galaxies from star-forming galaxies and not to confuse dusty star-formers as quiescents \citep[e.g.,][]{2005ApJ...624L..81L,Williams2009,2009ApJ...706..885W,Whitaker2012,2014MNRAS.440.1880W}. 
The UVJ color-color selection of passive galaxies is particularly efficient at $z=1-3$, where the 4000\AA~break is moving through the medium-band filters.
\citet{straat}, show an in-depth analysis of the UVJ color-color selection, as well as confirming passive galaxies at $z=3$.
Using rest-frame colors from NMBS, we construct a UVJ color-color diagram for our sample, shown in Figure~\ref{fig:uvj}.
Galaxies that lie above the relation defined by (U~$-$~V)~$\textgreater0.87~\times$~(V~$-$~J)~$+~0.60$, (U~$-$~V)~$\textgreater1.3$, and (V~$-$~J)~$\textless1.6$) are considered to be quiescent.
The total number of star-forming and quiescent galaxies in each sample can be seen in Table~\ref{table:sam}.

The fraction of star-forming and quiescent galaxies as a function of cluster radius is shown in Figure~\ref{fig:rfrac}.
We calculate the red/blue fractions in our three radius intervals, {\radin}, {\rad}, and {\Rv}$\textgreater2~$. 
The error for each fraction is estimated by assuming a beta distribution following \citet{2011PASA...28..128C}. 
In the field, the fraction of quiescent galaxies is $33\pm{1}\%$ compared to star-forming galaxies at $67\pm{1}\%$.
However, the fraction of quiescent galaxies increases to $55\pm{5}\%$ within 0.5~{\Rv}.
The increase of quiescent galaxy fraction as a function of clustocentric distance is suggestive that the density-SFR relation is in place at $z\sim1$.
However, we are defining environment using {\Rv} and not surface-density, therefore we cannot explicitly trace the changes in environmental density.

\section{Analysis}
\label{sec:ana}
\subsection{Determination of Structural Parameters}
We use the CANDELS \citep{2011ApJS..197...35G,2011ApJS..197...36K} {\HST}/ACS F814W ($\lambda\sim0.42$ $\mu$m rest-frame) image that contains our field and cluster galaxy samples to measure galaxy sizes. 
The pixel scale of this image is $0.03''$/pixel.
We use {\gal} \citep{2010ApJ...721..193P} to measure the half-light radii of the semi-major axis ($r_{1/2,maj}$) of each galaxy based on a single {\n} light profile.
{\gal} is run in batch mode, using a python wrapper, on a masked, background subtracted image of each galaxy. 
Additional inputs for {\gal} include: a point-spread-function (PSF) image, a sigma image, and a constraint file.
The constraint file only limits the bounds for the {\n} index to $0.2-8$, {\n} values above this limit indicate a poor fit \citep[e.g.,][]{Raichoor2012}.
We explain the process for creating each input below.

Individual galaxy images are created by cutting $90\times90$ pixels, or $\sim 20 \times 20$ kpc at {\z}, thumbnails from the {\HST}/ACS F814W image.
Each thumbnail has a mask that flags all objects outside of $1.2''$ from the image centre.
We do not mask inside of $1.2''$ because masking close neighbours to the central galaxy may mask some of the central galaxy's light.
Instead, we allow {\gal} to do a multi object fit inside this radius.
The masking is accomplished by using {\sx} with a detection threshold of $2.5\sigma$ above the background rms level to create a bad pixel mask.  

We create a sigma image that has a constant value of flux equal to the standard deviation of the flux distribution in the region around the object.
The size of the annulus that the flux is measured in has a diameter of $0.6''$.
The background in each image is estimated from {\sx} and then subtracted from the image using {\iraf}'s IMARITH package.

The use of an accurate PSF is crucial for measuring reliable sizes.
We use the PSF image created by the 3DHST survey team \citep{2014ApJS..214...24S}.
For a full description of the construction of the PSF see section 3.3 (and appendix A) of their paper.
The FWHM of the PSF is $0.11''$ ($\sim0.9$ kpc at {\z}); we can reliably measure sizes down to FWHM/2, $\sim0.5$ kpc \citep{vanDokkum2010,2015ApJ...808L..29S}.

After running {\gal}, we exclude galaxies that may have unreliable measured sizes.
Measured galaxy sizes from {\gal} can be unreliable if one or more of the galaxy's structural parameters is equal to the boundary value given in the constraint file or if they are flagged by {\gal} \citep[see ][]{2012ApJS..203...24V}.
We also remove galaxies if they are unresolved in the NMBS ground based image, but are resolved multicomponent systems in the {\HST} image.
The fractions of field star-forming and quiescent galaxies that remain after removing galaxies with unreliable sizes are 1189/1397 ($85\%$ complete) and 535/690 ($78\%$ complete), respectively.
For the 2~{\Rv} cluster star-forming and quiescent galaxies the remaining galaxy fractions are 176/206 ($85\%$ complete) and 91/97 ($94\%$ complete), respectively.
The final fractions of 0.5~{\Rv} cluster star-forming and quiescent galaxies are 34/38 ($89\%$ complete) and 44/47($94\%$ complete), respectively.
We investigate any possible magnitude or mass dependence of galaxies that fail the GALFIT fitting procedure and find no dependence on mass or magnitude, therefore, these failed fits do not affect our results.
\begin{figure}
\epsscale{1.18}\plotone{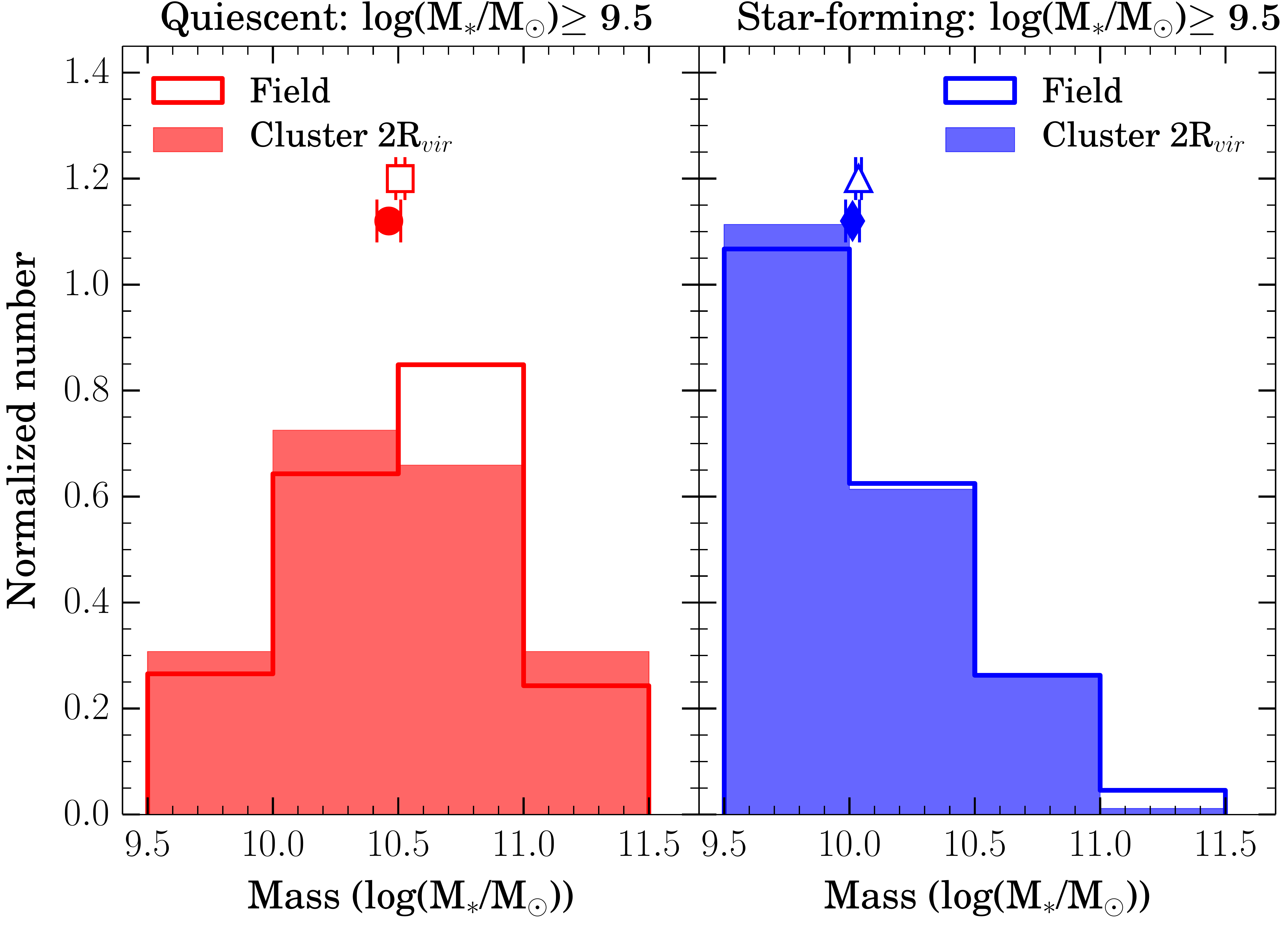}\\
\epsscale{1.18}\plotone{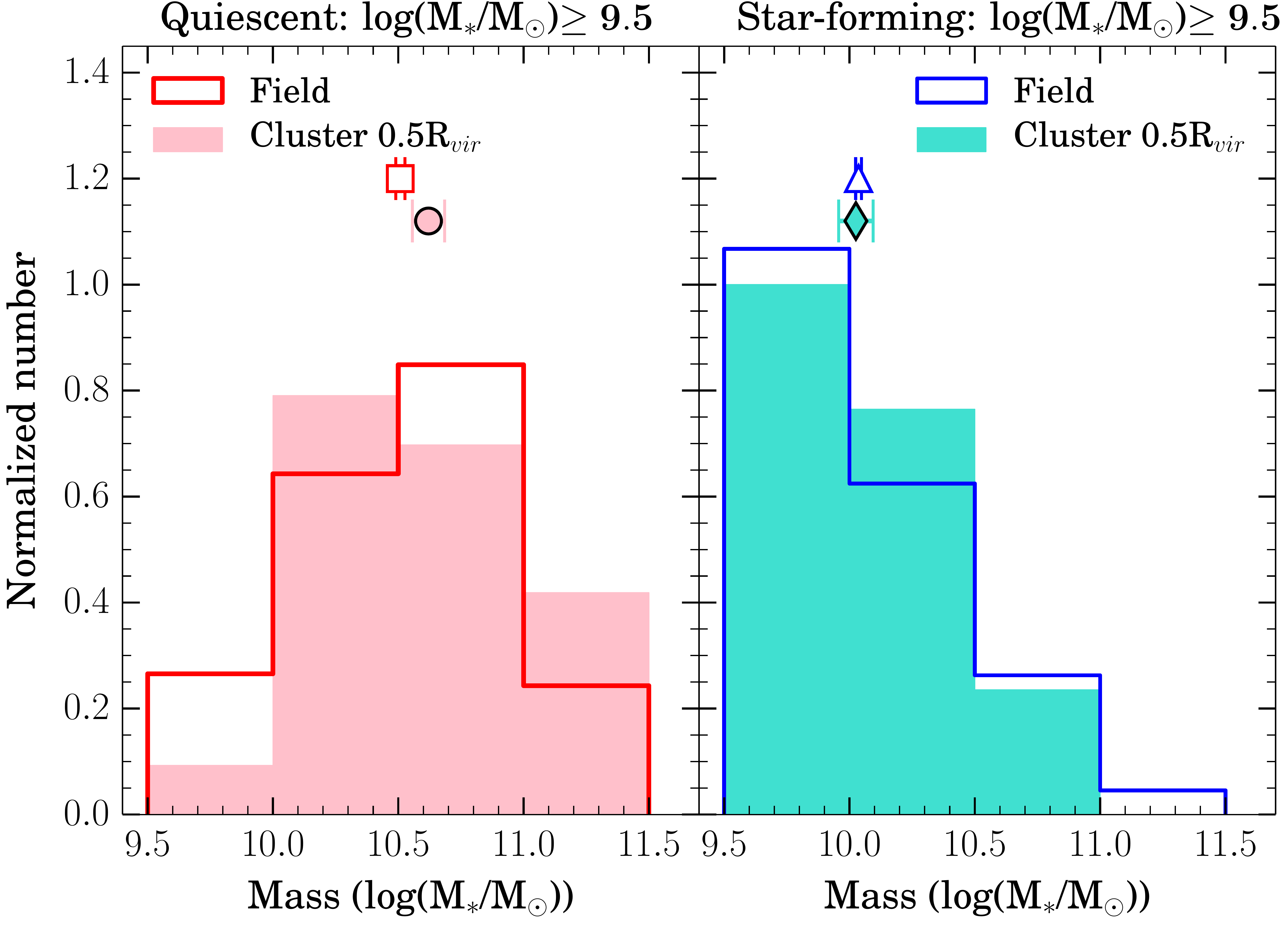}
  \caption{Top: Area-normalised mass distributions for quiescent (left) and star-forming (right) field (open histograms) and 2~{\Rv} cluster galaxies (closed histograms). The mass distributions and average masses for quiescent field (open squares) and cluster galaxies (filled circles) are similar and therefore differences in size should not be driven by differences in mass. The mass distributions and average masses for star-forming field (open triangle) and cluster galaxies (filled diamonds) are also consistent. In the bottom panels, area-normalised mass distributions and averages are shown for the 0.5~{\Rv} cluster sample. Again, no significant differences in the mass distributions are seen.}
  \label{fig:mass}
\end{figure}
\section{Results}
\label{sec:res}
\subsection{Mass-normalised Sizes}
In Figure~\ref{fig:mass}, we show the mass distributions for each subsample as well as the average mass.
We find that the distributions are similar across environment for the quiescent and star-forming samples.
In addition, we use a two-sample KS test to determine if the galaxies in each of the 4 sub-samples shown in Figure ~\ref{fig:mass}, are drawn from the same parent sample for each cluster radius and galaxy type. 
We find that for all sub-samples, field and cluster galaxies are consistent with being drawn from the same parent population all having P$\textless$0.14 or $\textless1.5\sigma$.
\begin{figure*}
\epsscale{2.1}\plottwo{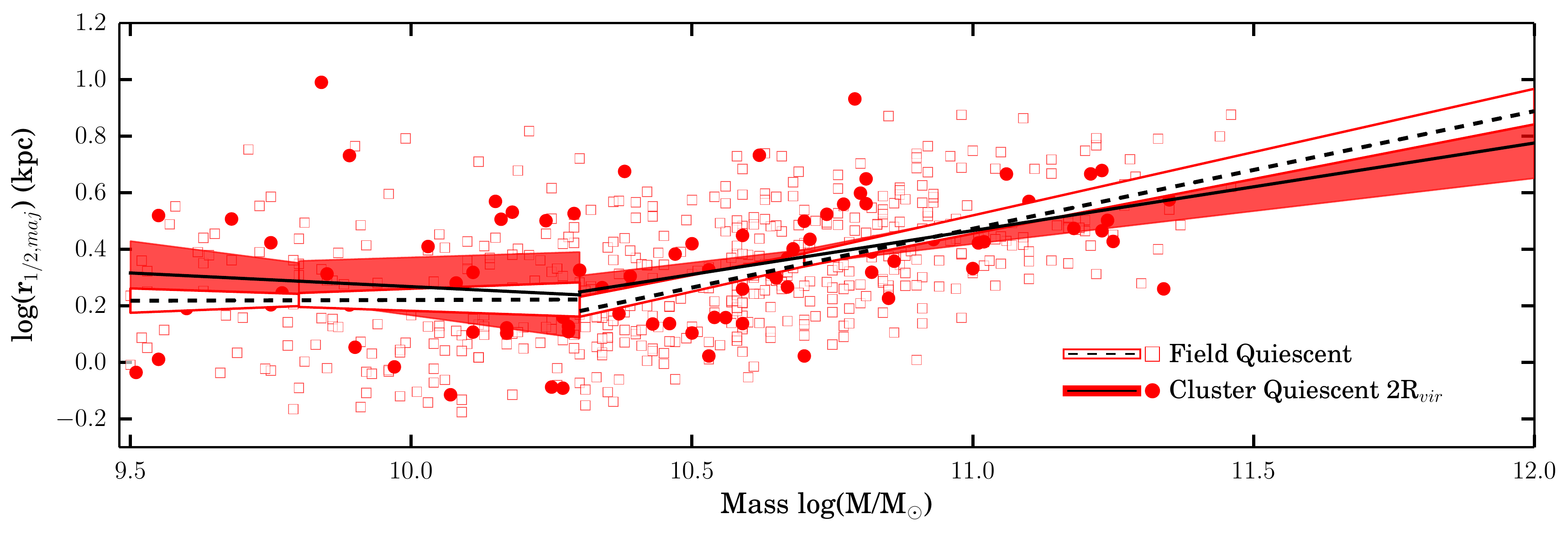}{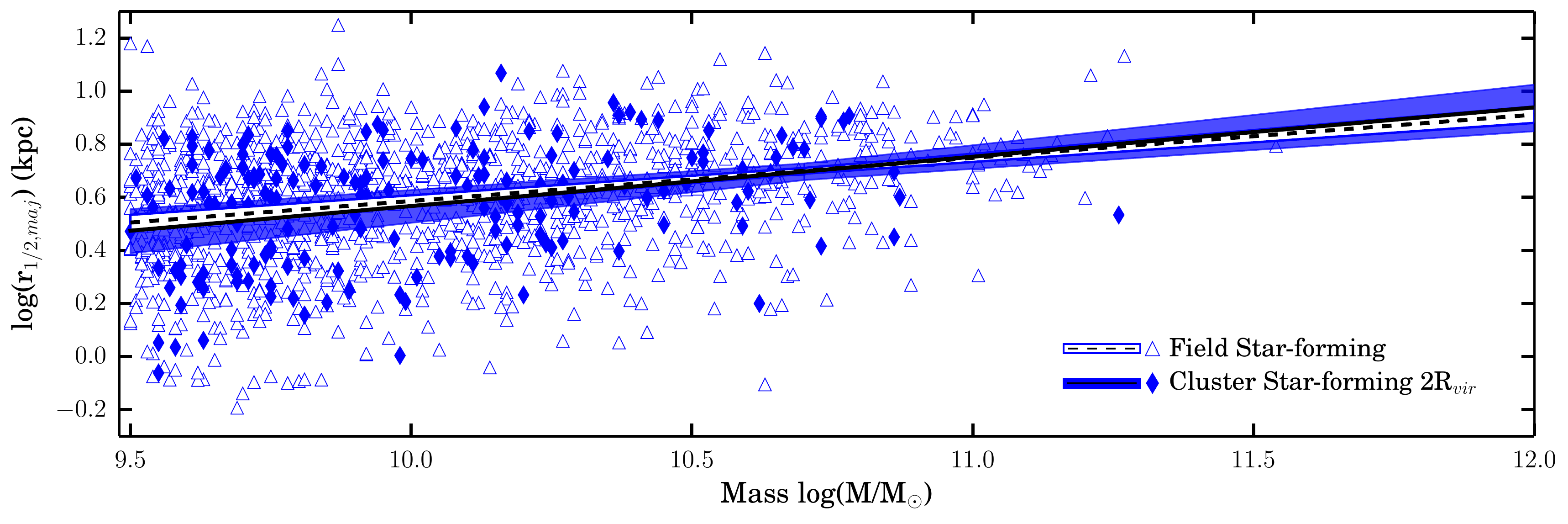}\\
\epsscale{2.1}\plottwo{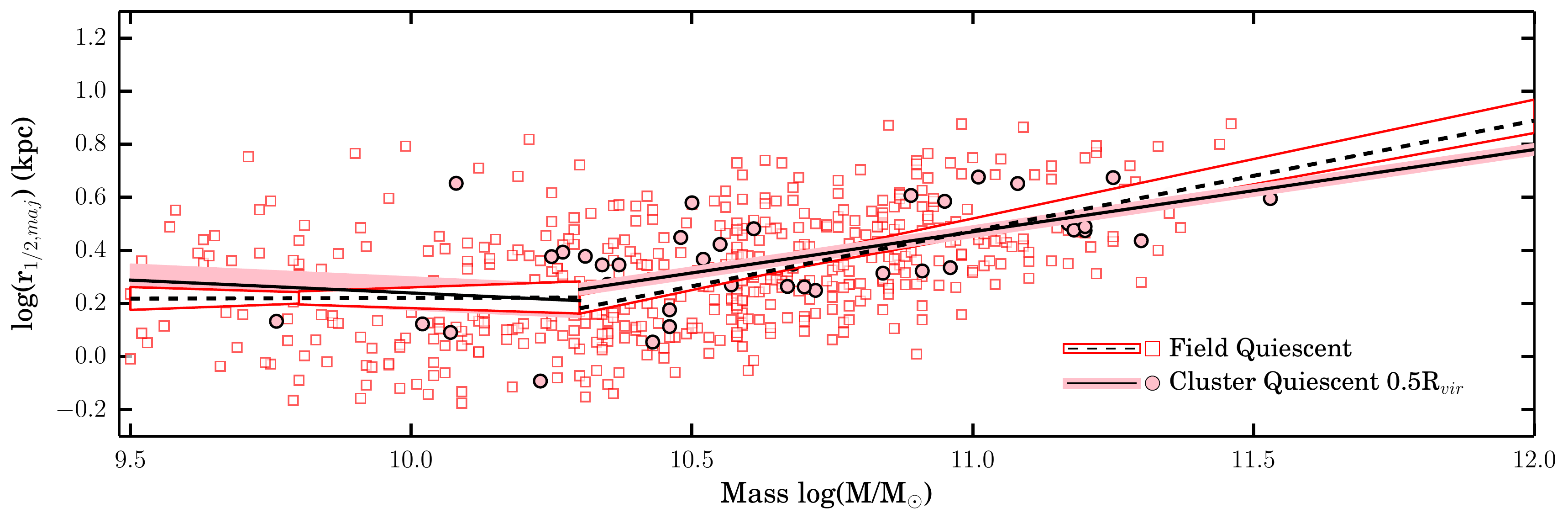}{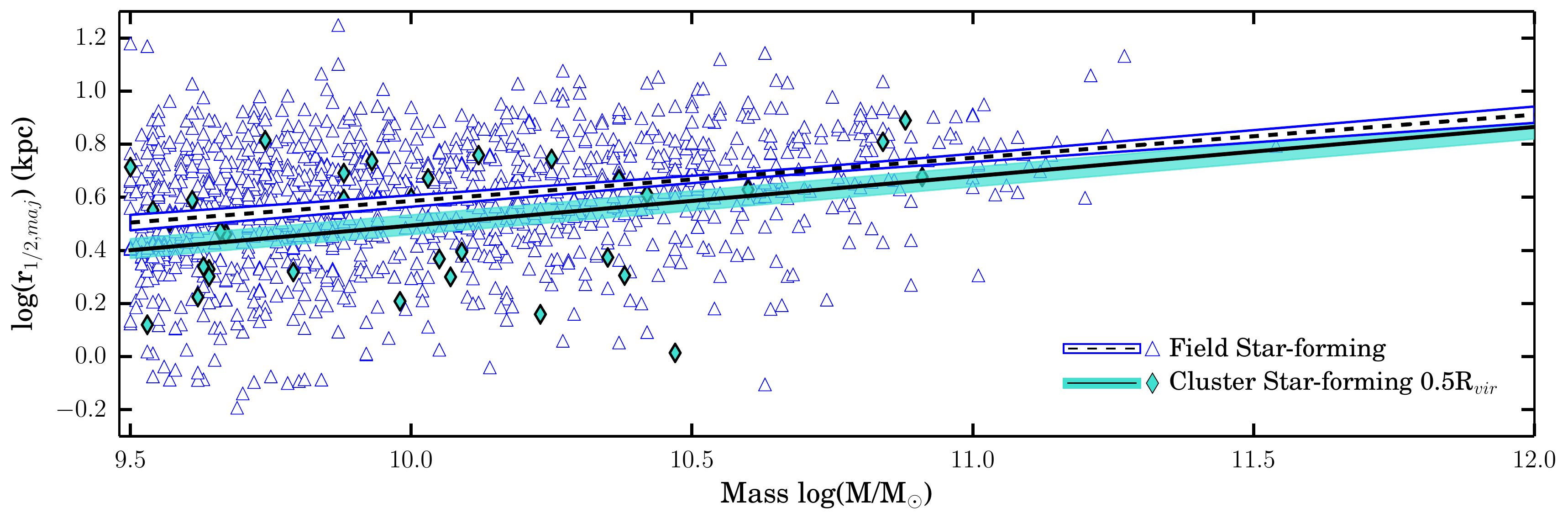}
  \caption{{\ms} distributions for quiescent and star-forming, field and cluster galaxies. In all panels, the field star-forming and quiescent galaxies are represented by open blue triangles and red squares, respectively. In the top two panels, the cluster samples are selected within 2~{\Rv} and are shown as solid blue diamonds (star-forming) or solid red squares (quiescent). The best-fits to the data are shown as solid (dashed) lines for cluster and field galaxies. The respective bootstrap errors for these fits are shown as filled (open) contours. The best-fits, their bootstrap errors, and average sizes are shown in Table~\ref{table:fit} and Table~\ref{table:gal}. The best-fit for field and 2~{\Rv} cluster quiescent galaxies with {\M}$\textgreater10.3$ are consistent. The best-fits are consistent for field and 2~{\Rv} cluster star-forming galaxies. The bottom two panels are the same, except the cluster samples are selected with 0.5~{\Rv} and the colors of the symbols are a lighter color. The slope is fixed to the best-fit value obtained from the 2~{\Rv} fits for each environment and galaxy type. For cluster galaxies within 0.5~{\Rv}, the mass$-$normalised average size for field and cluster quiescent galaxies remain roughly consistent. However, the mass$-$normalised average size for star-forming field galaxies is $16\pm{7}\%$ larger than that of star-forming 0.5~{\Rv} cluster galaxies.}
  \label{fig:size}
\end{figure*}
\begin{table*}
\begin{center}
\caption{Best-fit values for $A$ and $\alpha$ to determine the mass$-$normalised average sizes of the form: \\ $r~(m_{*})/{kpc}=A~{\cdot}~{m}_{*}^{\alpha}$, where $m_{*}~{\equiv}~M_{*}/5\times10^{10}~$M$_{\odot}$}
\label{table:fit}
	\begin{tabular}{c c c c c}
	\hline\hline
	&Quiescent$^1$&&Star-forming  \\[0.5ex]
	\hline
	Environment & $\phantom{}$log($A$)$\phantom{}$& $\phantom{}$$\alpha$$\phantom{}$& $\phantom{}$log($A$)$\phantom{}$&$\alpha$$\phantom{}$\\[0.5ex]
	\hline\\[-1.5ex]
	Field & $0.35\pm{0.01}$ & $0.42\pm{0.03}$ & $0.70\pm{0.01}$ & $0.16\pm{0.02}$\\[0.5ex]
	Cluster (2~{\Rv}) & $0.37\pm{0.03}$ & $0.31\pm{0.08}$ & $0.70\pm{0.03}$ & $0.19\pm{0.04}$\\[0.5ex]
	Cluster (0.5~{\Rv})$^{2}$ & $0.38\pm{0.02}$ &$-$& $0.62\pm{0.04}$ &$-$\\[0.5ex]
	\hline \\
	\end{tabular}
	\end{center}
	\vglue -5ex
			$\phantom{0000000000000000000000000000000}$$^{1}$ Fits are only shown for galaxies with {\M}$\textgreater10.3$\\
			$\phantom{0000000000000000000000000000000}$$^{2}$ $\alpha$ is fixed to the 2~{\Rv} value
\end{table*}
\begin{table*}
\begin{center}
\caption{{\n} indices and mass$-$normalised average sizes of Star-forming and Quiescent \\ Field and Cluster Galaxies derived from {\HST}/ACS F814W images}
\label{table:gal}
	\begin{tabular}{c c c c c c}
	\hline\hline
	&Quiescent$^{1}$&&Star-forming  \\[0.5ex]
	\hline
	Environment & $\phantom{}$$r_{1/2, maj}$ $\phantom{}$& $\phantom{}n$ $\phantom{}$& $\phantom{}$$r_{1/2, maj}$ $\phantom{}$& $n$$\phantom{}$\\[0.5ex]
	&$\phantom{}$(kpc) $\phantom{}$&& $\phantom{}$(kpc) $\phantom{}$&\\[0.25ex]
	\hline\\[-1.5ex]
	Field &$\phantom{}2.23\pm{0.04}$$\phantom{}$& $\phantom{}1.96\pm{0.03}$ $\phantom{}$& $5.01\pm{0.13}$ $\phantom{}$& $0.94\pm{0.02}$ $\phantom{}$\\[0.5ex]
	Cluster (2~{\Rv}) & $\phantom{}2.36\pm{0.13}$ $\phantom{}$& $\phantom{}1.96\pm{0.07}$ $\phantom{}$& $\phantom{}4.98\pm{0.37}$ $\phantom{}$& $0.92\pm{0.04}$ $\phantom{}$\\[0.5ex]
	Cluster (0.5~{\Rv}) & $\phantom{}2.38\pm{0.11}$ $\phantom{}$& $\phantom{}2.05\pm{0.07}$ $\phantom{}$& $\phantom{}4.20\pm{0.36}$ $\phantom{}$& $1.22\pm{0.11}$ $\phantom{}$\\[0.5ex]
	\\[-2.5ex]
	$\Delta_{FC}$$^{2}$ (2~{\Rv})&$-0.13\pm{0.14}$ ($0.9\sigma$)&$0.0\pm{0.08}$ ($0\sigma$)& $0.03\pm{0.39}$ ($0.08\sigma$)&$\phantom{}$$-0.02~\pm~{0.04}$~($0.22\sigma$)\\
	\\[-2.5ex]
	$\Delta_{FC}$ (0.5~{\Rv})&$-0.15\pm{0.12}$ ($1.25\sigma$)&$-0.09\pm{0.08}$ ($1.12\sigma$)& $0.81\pm{0.38}$ ($2.20\sigma$)&$\phantom{}$$-0.28~\pm~{0.11}$~($2.5\sigma$)\\
	\hline \\
	\end{tabular}
	\end{center}
	\vglue -5ex
			$\phantom{00000000000000000}$$^{1}$ Sizes and {\n} indices are for galaxies with {\M}$\textgreater10.3$ \\
			$\phantom{00000000000000000}$$^{2}$ $\Delta_{FC}\equiv$ Field~$-$~Cluster
\end{table*}

To probe the size differences of field and cluster galaxies, we compare their mass$-$normalised average sizes \citep[see,][for further details]{2015ApJ...806....3A}.
We fit for average size over the entire size distribution of galaxies using a mass normalisation instead of binning by mass.
If field galaxies are not mass$-$matched to the cluster galaxy sample, then measuring the average size in mass bins can lead to biased results. 
We fit for the average size as a function of mass using the parametrisation:
\begin{equation} r~(m_{*})/\text{kpc}=A~{\cdot}~\text{m}_{*}^{\alpha} \end{equation}

For the comparison of the 2~{\Rv} cluster and field galaxy sizes we compute the best fit for both the slope, $\alpha$, and y-intercept, $A$, of the mass-size relation.
Where $m_{*}$ is the ratio of the galaxy stellar mass to a constant mass defined below.
Errors in the average size and slope are determined from bootstrapping the fit for $A$ and $\alpha$.
For the 0.5~{\Rv} cluster and field galaxy average sizes we compute the best fit for $A$ only, and fix $\alpha$ to the value obtained in the fit for the 2~{\Rv} cluster and field average sizes.
The error in $A$ is then obtained from bootstrapping.

In Figure~\ref{fig:size}, we show the {\ms} distributions for our sample of quiescent and star-forming, field and cluster galaxies.
We show the {\ms} relation for both the 2~{\Rv} and 0.5~{\Rv} cluster samples (top and bottom panels, respectively).
The best-fits and their errors are shown as lines and contours in Figure~\ref{fig:size} and are listed in Table~\ref{table:fit}.

For quiescent galaxies, the {\ms} relation flattens at low masses, and can be seen in Figure~\ref{fig:size} at {\M}$\leq10.3$.
The cause of this flattening may be due to a difference in the projected axis ratios of high and low mass quiescent systems \citep[e.g.,][]{2013ApJ...773..149C}.
We use the same mass cut as \citet{2014ApJ...788...28V}, and fit quiescent galaxies with {\M}$\textless10.3$ and {\M}$\textgreater10.3$ separately, using different values for $m_{*}$.
Field and cluster quiescent galaxies with {\M}$\leq10.3$, are fit using $m_{*}~{\equiv}~M_{*}/6\times10^{9}~$M$_{\odot}$.
We fit field and cluster galaxies with {\M}$\textgreater10.3$ using $m_{*}~{\equiv}~M_{*}/5\times10^{10}~$M$_{\odot}$.

We find that field and 2~{\Rv} cluster quiescent galaxies with {\M}$\leq10.3$ have consistent mass$-$normalised average sizes.
Field and 2~{\Rv} cluster quiescent galaxies with {\M}$\textgreater10.3$ do not have a significant difference in their average sizes, $\Delta_{FC}=-0.14\pm{0.11}$ ($1.27\sigma$), and we can rule out any size difference greater than $6\%$.
The best-fits for $A$ and $\alpha$ are listed in Table~\ref{table:fit}, and the mass$-$normalised sizes that correspond to these fits are shown in Table~\ref{table:gal}.

The mass$-$normalised average sizes for field and 0.5~{\Rv} cluster quiescent galaxies remain consistent at masses above and below {\M}$=10.3$.
The lack of any significant size difference between field and cluster quiescent galaxies, regardless of cluster distance, indicates that environment is not accelerating their size growth.

We fit star-forming galaxies with {\M}$\textgreater9.5$ using $m_{*}~{\equiv}~M_{*}/5\times10^{10}~$M$_{\odot}$.
Field and 2~{\Rv} cluster star-forming galaxies have consistent mass$-$normalised average sizes, $\Delta_{FC}=0.03\pm{0.39}$ ($0.08\sigma$).
However, the average size for field star-forming galaxies is $16\pm{7}\%$ larger than that of 0.5~{\Rv} cluster star-forming galaxies, $\Delta_{FC}=0.81\pm{0.38}$ ($2.20\sigma$).
This result remains even if we do not fix the slope.
The smaller sizes we find for 0.5~{\Rv} star-forming galaxies suggests that environment is either acting on their growth mechanisms, i.e., quenching their SFRs and stunting their growth, or disrupting their stellar disks and truncating their light profiles.

Different from some studies, we have selected our sample using UVJ colors. 
We now test if selecting by {\n} index/morphology \citep[e.g.,][]{2004MNRAS.353..713K,2014ApJ...788...51N, 2015MNRAS.450.1246K} affects our results. 
We reselect our samples of galaxies with $n$ above and below $n=2.5$, and refit the sizes. 
We find that there is no significant change in our results.

\subsection{{\n} indices}
\begin{figure}[hb]
\epsscale{1.18}\plotone{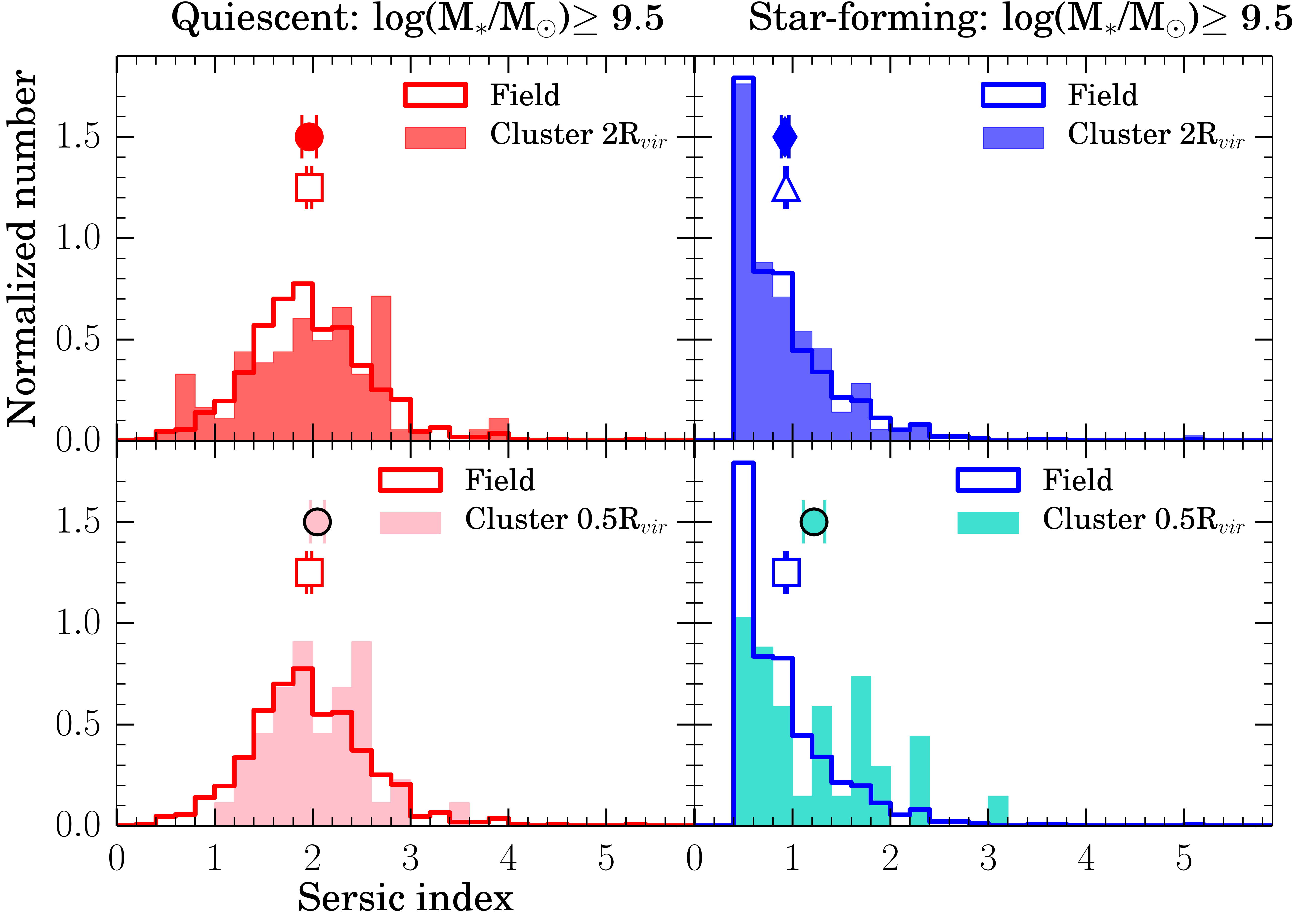}
  \caption{Area-normalised distributions of the {\n} indices of quiescent (left) and star-forming (right) field (open histograms) and cluster (closed histograms) galaxies. The top panels are for the cluster samples drawn within apertures of 2~{\Rv}. The distributions of {\n} index for field and cluster quiescent galaxies are consistent. The same is true for star-forming galaxies. The bottom two panels are the same except the cluster sample aperture is 0.5~{\Rv}. Quiescent galaxies remain consistent in distributions of $n$, however, the fraction of star-forming galaxies with $n\textgreater1$ is $0.5$ in the cluster core compared to $0.31$ for field star-forming galaxies.}
  \label{fig:n}
\end{figure}

In Figure~\ref{fig:n}, we show the area-normalised distribution of {\n} indices and their averages for field and cluster, star-forming and quiescent galaxies obtained from {\gal}.
Errors in the average values are estimated using the error in the mean.

In Table~\ref{table:gal}, we show these averages and their errors as well as the significance of their difference between field and cluster.
The top panels of Figure~\ref{fig:n} are for 2~{\Rv} cluster galaxies while the bottom panels are for 0.5~{\Rv} cluster galaxies.
The distribution of {\n} indices is similar for both field and cluster quiescent galaxies, regardless of cluster-centric radius.
The fraction of quiescent field galaxies with $n\textgreater2$ is $0.42$, compared to $0.52$ for the cluster outskirts, and $0.49$ for the cluster core.
Where the average {\n} index for field quiescent galaxies is $1.96\pm{0.03}$ compared to the cluster outskirts, $n=1.96\pm{0.07}$, and cluster core, $n=2.05\pm{0.07}$.
The lack of a difference in the distributions or averages in {\n} index for field and cluster quiescent galaxies, regardless of cluster distance, suggests that these galaxies are most likely similar in morphologies, and we cannot use {\n} index to differentiate their growth mechanisms. 
\begin{figure*}
\epsscale{1.18}\plotone{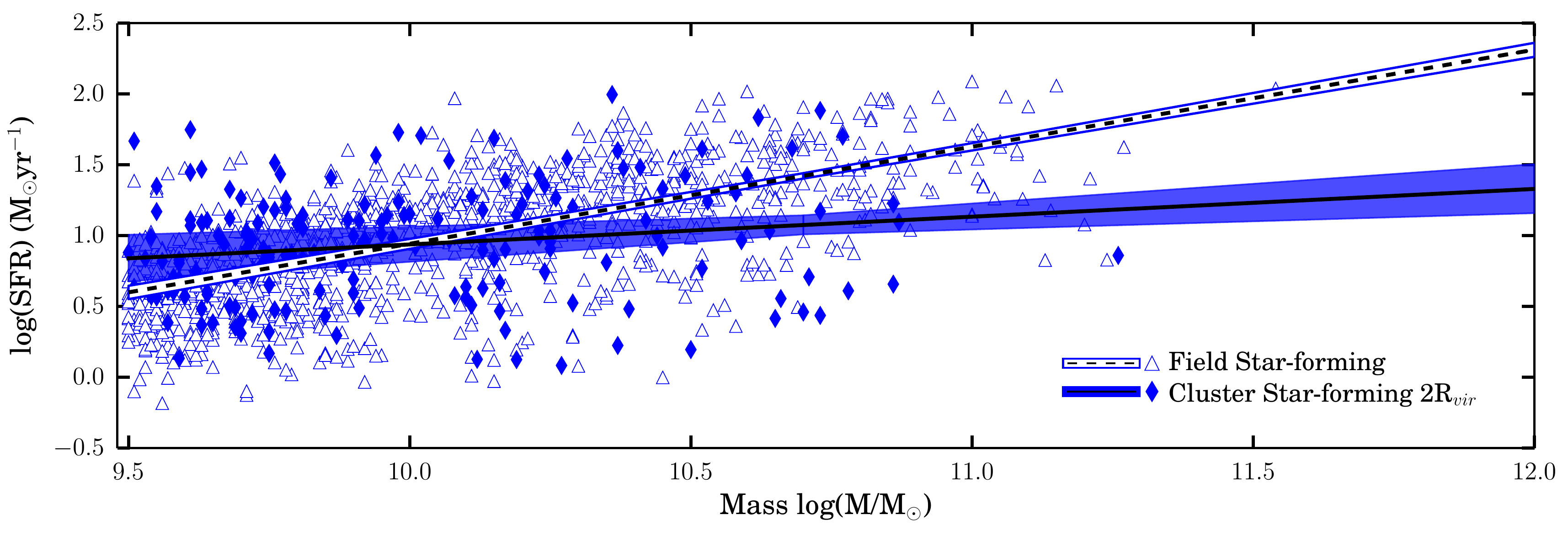}\\
\epsscale{1.18}\plotone{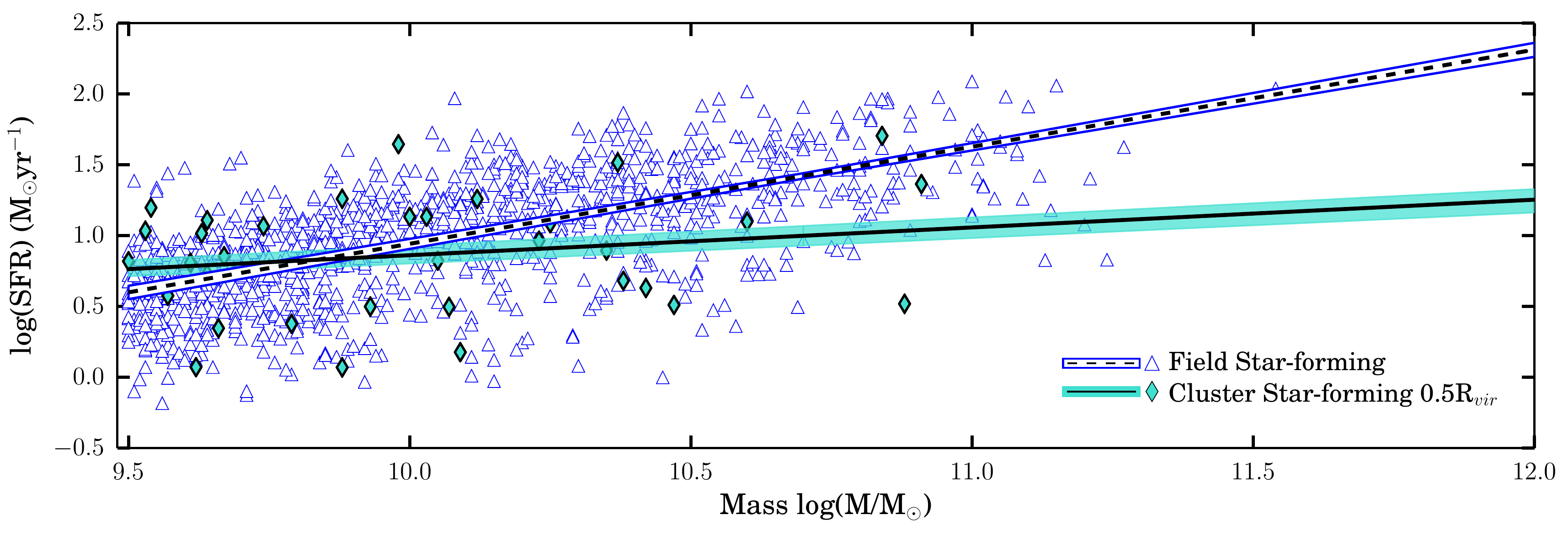}
  \caption{Mass$-$SFR distributions for star-forming, field and cluster galaxies. In both panels, the field star-forming galaxies are represented by open blue triangles. In the top panel, the cluster samples are selected within 2~{\Rv} and are shown as solid blue diamonds. In the bottom panel, we show cluster galaxies chosen within 0.5~{\Rv} as turquoise diamonds. The best-fits to the mass$-$SFR distributions are shown as solid (dashed) lines for cluster and field galaxies. The respective bootstrap errors for these fits are shown as filled (open) contours. The derived average sizes and their errors are shown in Table~\ref{table:sfr}. The best-fits for the field and cluster galaxies are different by $\sim8\sigma$, indicating that the cluster galaxies are undergoing quenching. }
    \label{fig:sfr2}
\end{figure*}
\begin{figure}
\epsscale{1.18}\plotone{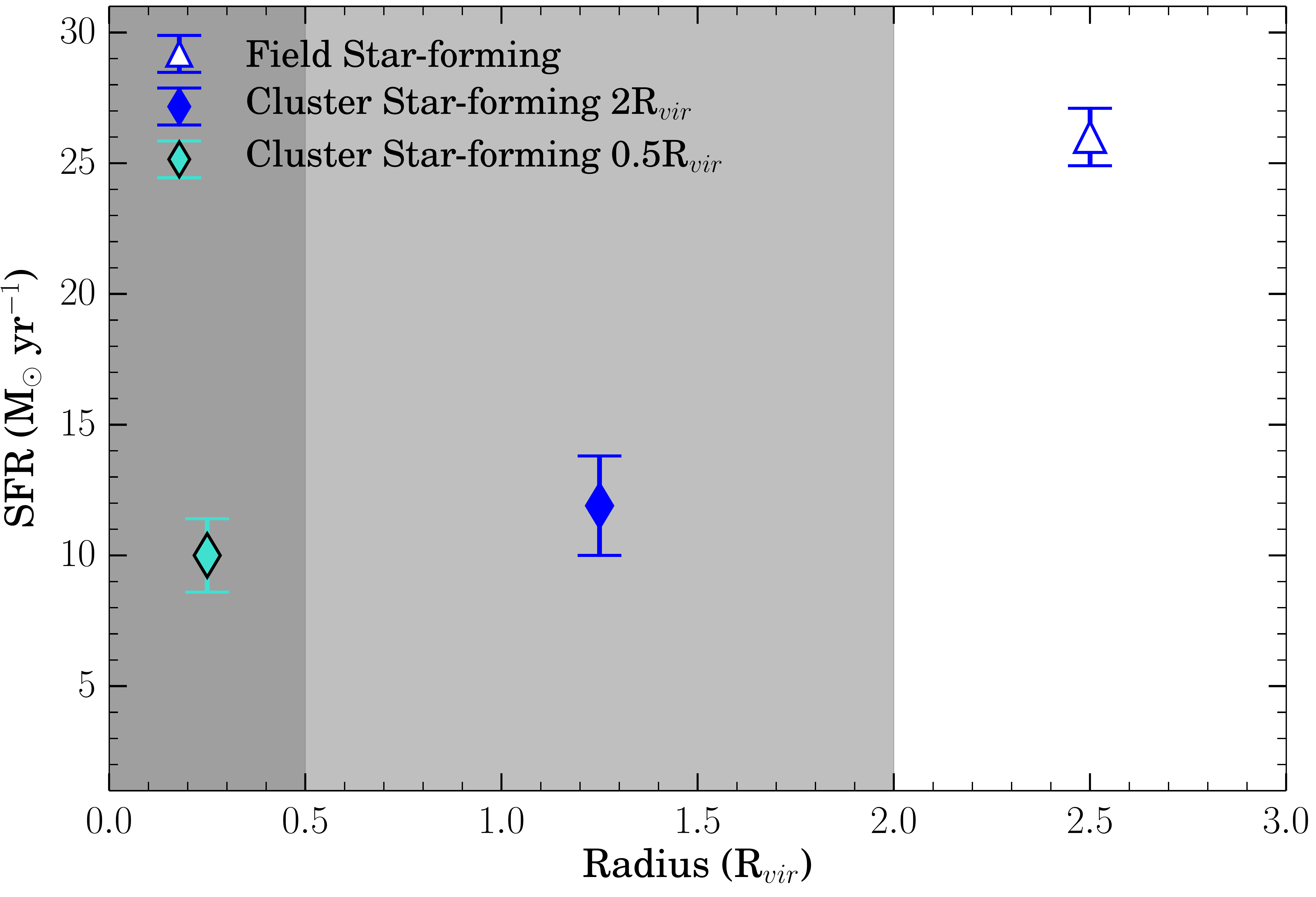}
  \caption{Mass-normalised average star-formation rates (SFR) for field and cluster star-forming galaxies as a function of {\Rv}. The averages are measured within three cluster-centric distances of $0-0.5$~{\Rv}, $0.5-2$~{\Rv}, and $\textgreater2~${\Rv}. Errors are calculated from bootstrapping the fit to the mass$-$SFR relation from Figure~\ref{fig:sfr2}. The average SFR of field star-forming galaxies is up to $2.5$ times larger than that of the cluster. This indicates that the environment is likely quenching galaxies.}
    \label{fig:sfr}
\end{figure}

The distributions of {\n} indices and the average {\n} index of field and cluster star-forming galaxies at 2~{\Rv} are consistent. 
We do note, however, that the 0.5~{\Rv} cluster galaxies have a different distribution of $n$ relative to the field galaxies. 
Galaxies within 0.5~{\Rv} have an equal fraction of $n\textgreater1$ and $n\textless1$ {\n} indices. 
This is in stark contrast to the field, where the fraction of galaxies with $n\textless1$ is 0.69, see Figure~\ref{fig:n}.
The distribution in {\n} indices of the cluster core galaxies could be indicative of a distribution in spectral types, as observed in \citet{2013ApJ...777..124A}, while the field star-forming population may be more uniform.

\subsection{Star-Formation Rates}
By comparing the SFRs of field and cluster, star-forming galaxies we can quantify possible differential growth.
We use the SFRs from the NMBS UV$+$IR SFR catalog for individual galaxies.
For more details on the derivation of SFRs, please see \citet{2012ApJ...754L..29W}.

Because there is a known trend of increasing SFR with mass \citep[e.g.,][]{2004MNRAS.351.1151B,2010ApJ...721..193P}, we compare SFRs between field and cluster by determining mass-normalised average SFRs.
We fit the mass$-$SFR relations using the same parameterisation used to fit our {\ms} relations.
Again, we use $m_{*}~{\equiv}~M_{*}/5\times10^{10}~$M$_{\odot}$ for galaxies with {\M}$\textgreater9.5$.

In Figure~\ref{fig:sfr2}, we show the mass$-$SFR distributions and their best-fits for our samples of star-forming galaxies.
The error in the average SFR is derived by bootstrapping the fit, and can be seen in Figure~\ref{fig:sfr2} as filled and open contours.
The average SFRs, their errors, and the significance at which they differ from each other can be seen in Table~\ref{table:sfr}.
Field galaxies have an average SFR of $26.4\pm{1.1}$~{\y} compared to $11.9\pm{1.9}$~{\y} for the cluster outskirts and $10.0\pm{1.4}$~{\y} for the cluster core.
While this difference has a significance of $7-9\sigma$, it can be seen in Figure~\ref{fig:sfr}, that the larger difference in SFR between field and cluster occurs for galaxies with {\M}$\textgreater10.3$.
Therefore, it is likely that the higher mass galaxies drive the difference in SFRs, and that the cluster environment is effective at suppressing or quenching the star-formation of massive galaxies, while lower mass cluster and field galaxies have similar SFRs.

In Figure~\ref{fig:sfr}, we show the mass-normalised average SFRs as a function of clustocentric radius.
The average SFR drops by a factor of two within 2~{\Rv}, however, the the average SFRs between the cluster outskirts and core are consistent, suggesting that environmental effects extend to 2~{\Rv}.
\section{Discussion}
\label{sec:dis}
We have used a sample of star-forming and quiescent, field and cluster galaxies at {\z} to study the influence of environment on galaxy properties.
For the first time, we use different values of {\Rv} to probe the effects of environment on galaxy sizes.
It appears that only the star-forming galaxies found within 0.5~{\Rv} of the cluster centres show a significant difference in size and {\n} index compared to their field counter-parts at {\z}.
The difference of these two populations is further supported by the suppressed SFRs of cluster star-forming galaxies compared to field star-forming galaxies.
On the other hand, quiescent field and cluster galaxies have consistent sizes {\n} indices, and SFRs, independent of cluster-centric radius.
\begin{table}
\begin{center}
\caption{mass$-$normalised SFRs for field and cluster star-forming galaxies. SFRs have units of M$_{\odot}$ yr$^{-1}$.}
\label{table:sfr}
	\begin{tabular}{c c}
	\hline\hline
	Environment & SFR  \\[0.5ex]
	\hline\\[-1.5ex]
	Field & $26.4\pm{1.1}$  \\[0.5ex]
	Cluster (2~{\Rv}) & $11.9\pm{1.9}$  \\[0.5ex]
	Cluster (0.5~{\Rv}) & $10.0\pm{1.4}$  \\[0.5ex]
	\\[-2.5ex]
	$\Delta_{FC}$$^{1}$ (2~{\Rv})& $14.5\pm{2.2}$ ($6.6\sigma$)  \\
	\\[-2.5ex]
	$\Delta_{FC}$ (0.5~{\Rv})& $16.4\pm{1.8}$ ($9.2\sigma$)  \\
	\hline \\
	\end{tabular}
	\end{center}
	\vglue -5ex
			$\phantom{00000000000000000000000}$$^{1}$ $\Delta_{FC}\equiv$ Field~$-$~Cluster

\end{table}

\subsection{Quiescent Galaxies}
We do not find any significant difference between the sizes and {\n} indices of field and cluster quiescent galaxies; therefore, we can infer that they are evolving similarly.
\begin{figure}[]
\epsscale{1.17}\plotone{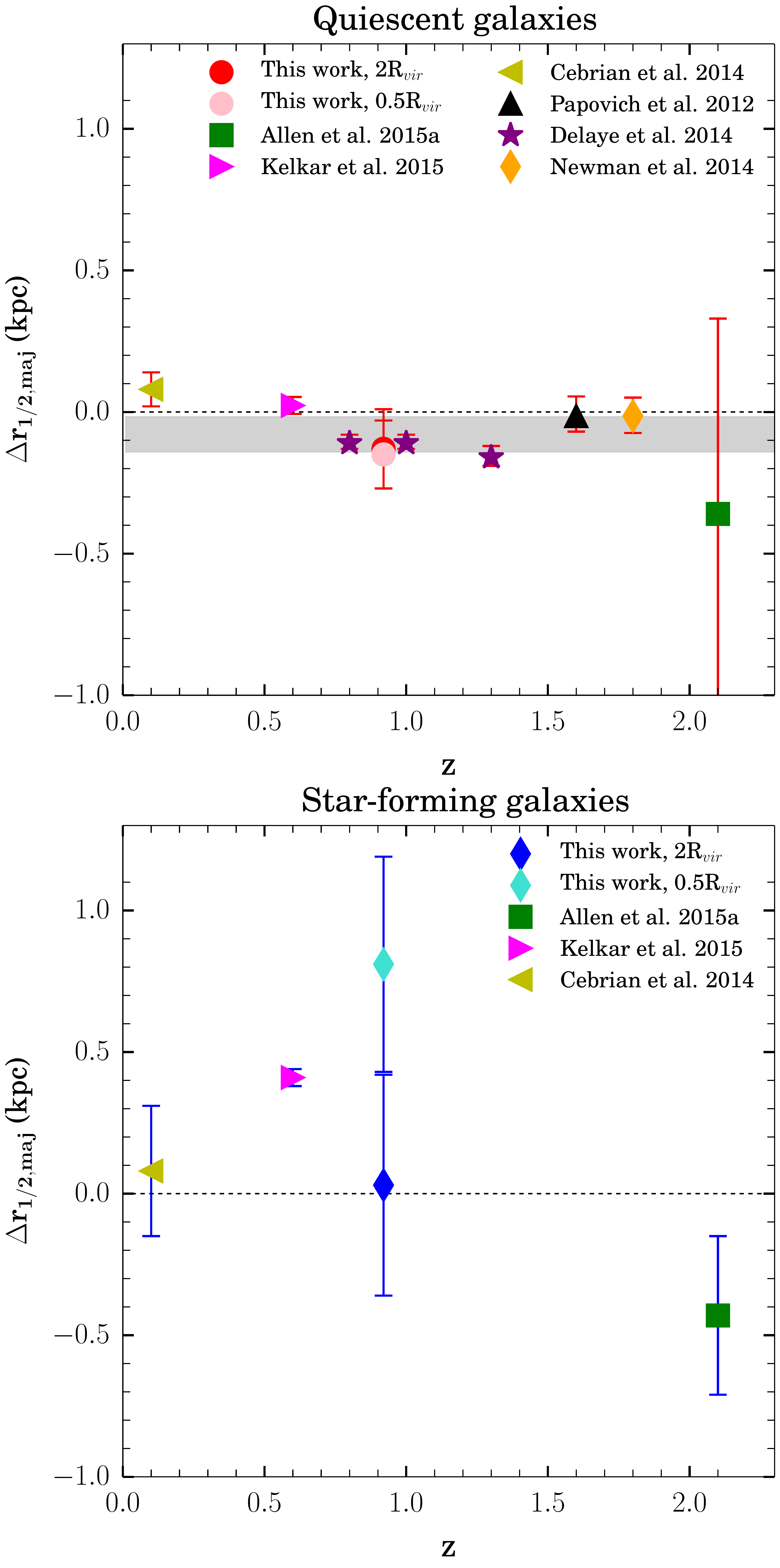}
  \caption{The evolution of $\Delta$r$_{1/2, maj}$ between the field and cluster for quiescent galaxies (top panel) and star-forming galaxies (bottom panel). $\Delta$r$_{1/2, maj}$ is the difference of r$_{1/2, maj}$ field minus r$_{1/2, maj}$ cluster. Our data are shown as red and pink filled circles (quiescent galaxies, 0.5~{\Rv} and 2~{\Rv}, respectively) and blue and turquoise filled diamonds (star-forming galaxies, 0.5~{\Rv} and 2~{\Rv}, respectively). We show a linear fit to the data, including one sigma errors, as a grey contour (the slope is fixed to zero) for quiescent galaxies. We weight the contribution of each point to the fit by its errors. The difference we measure for quiescent galaxies is marginal, $\Delta$r$_{1/2, maj}=-0.09\pm{0.06}$.}
  \label{fig:del}
\end{figure}
This result is consistent with other studies regardless of sample selection or redshift \citep[e.g.,][]{2013ApJ...779...29H,2014ApJ...788...51N,2015ApJ...806....3A,2015MNRAS.450.1246K}.
To illustrate this, we plot the difference in r$_{1/2, maj}$ between field and cluster quiescent galaxies at several redshifts, see Figure~\ref{fig:del}.
We adopted this approach from \citet{2014ApJ...788...51N} and use their data and as well as the work of this study, \citet{2015ApJ...806....3A}, \citet{2014MNRAS.444..682C}, and \citet{2015MNRAS.450.1246K}.

When we assume zero slope and preform a linear fit to the data we find an average difference of $-0.09\pm{0.06}$ kpc ($\textless2~\sigma$) between the sizes of field and cluster quiescent galaxies.
For an average quiescent cluster galaxy with {\M}=10.7 and r$_{1/2, maj}=2.4$ kpc, this difference would only represent a size offset of $4\%$ from a field quiescent galaxy with the same mass.
We conclude that environment does not affect the size evolution of quiescent galaxies.

The two data points from \citet{2014MNRAS.441..203D} deviate from zero size difference by a significant amount (see Figure~\ref{fig:del}).
The halo masses of the clusters in their study are on the order of $M_h\sim10^{15}$ M$_{\odot}$ where those in our work are on the order of $M_h\sim10^{13-14}$ M$_{\odot}$.
Larger sizes are expected for cluster galaxies that reside in larger mass halos \citep[e.g.,][]{2014MNRAS.439.3189S}, therefore, the lack of size difference measured in our work could be due to the lower mass halos of our clusters.

While it seems that there is little to no size difference between field and cluster quiescent galaxies, more studies are needed that consider the effects of different mass halos and different cluster-centric radii to constrain this apparent lack of environmental effects.

\subsection{Star-forming Galaxies}
We found no significant difference between the sizes and {\n} indices of star-forming galaxies in the outer cluster region ($2\textgreater${\Rv}$\textgreater0.5$) and field. 
However, within the cluster core ({\Rv}$\textless0.5$), star-forming galaxies have smaller sizes and {\n} indices with equal frequency above and below $n=1$, compared to the field.
Here, we explore what phenomena would cause environment to have a significant effect on star-forming galaxy sizes at {\Rv}$\textless0.5$, and star-forming galaxy SFRs within 2~{\Rv}.

The smaller sizes of the cluster core galaxies may be due to a combination of tidal stripping and harassment.
Therefore, a difference in dynamical timescales of galaxies that reside in the outer cluster regions versus the cluster core could explain the difference in their structural properties.
We use the velocity dispersions calculated in \citet{2014MNRAS.443.2679B} to estimate the different dynamical times, $t_{dyn}$, as a function of {\Rv} for galaxies in this study.
Galaxies in the outer cluster have $t_{dyn}\sim3.4$ Gyr compared to $t_{dyn}\sim1.7$ Gyr for galaxies in the cluster core.
Therefore, galaxies in the cluster core have most likely been exposed to environmental effects such as tidal stripping and harassment more frequently than those in the outer cluster.

Environmental effects would also cause suppressed star-formation that could be contributing to the smaller sizes (in rest-frame B-band).
This is consistent with our results where the average SFR of the core is a factor of 2.5 lower than the field.
However, the average SFR of the cluster outskirts is also suppressed compared to the field, but the average sizes between the outskirts and field are consistent.
This may be a result of a timescale issue where galaxies in the cluster outskirts have begun to quench, but that is not yet reflected in their sizes.
This is consistent with \citet{2013MNRAS.432..336W} who found that satellite galaxies remain star-forming for $2-4$ Gyr after their first cluster infall and then rapidly quench.
Furthermore, the difference in SFR between the field and cluster is likely driven by galaxies with {\M}$\textgreater10.3$, while the SFRs of lower mass galaxies are similar regardless of environment.
This is consistent with previous studies that find higher quenching efficiencies and lower specific SFRs for massive star-forming galaxies in groups \citep[e.g.,][]{2014ApJ...782...33L}.
Therefore, it is likely that environmental effects extend to 2~{\Rv}, however, the timescale for which a difference in size can be seen is longer than for the cluster core. 

There are few studies that compare the sizes of field and cluster star-forming galaxies at any redshift.
In Figure ~\ref{fig:del}, we show all of the current studies that measure the size difference for field and cluster star-forming galaxies.
From these few results it is unclear what role environment plays in the growth of star-forming galaxies.
To understand what is driving this size difference in the cores of clusters at {\z} and if it is occurring at lower or high redshifts, more studies that consider the effects of environment as a function of {\Rv} are needed.

While we chose to use clustocentric distance to define environment, it has been shown that surface density more strongly traces changes in galaxy populations in over-dense environments, such as spectral type and {\n} index \citep{2013ApJ...777..124A,2013ApJ...770...62D}.
To really understand how clusters affect their galaxies, it is important to understand how both environmental density and clustocentric radius play a role.

\section{conclusions}
\label{sec:con}
We have studied the dependence of the {\ms} relation on environment at different intervals of cluster {\Rv} using $\sim2400$ field and cluster galaxies at {\z}.
From the GEEC2 and NMBS surveys, we utilised accurate rest frame colors and stellar masses to select our mass-complete sample (down to {\M}$\geq9.5$) of star-forming and quiescent galaxies.
Our main results are as follows.\\

For quiescent galaxies:
\begin{itemize} 
  \item~We rule out a size difference of more than 6\%, regardless of the cluster-centric radius. Combining previous results from the literature, we determine that quiescent cluster galaxies are at most $0.09\pm{0.06}$ kpc larger in size than their field counter-parts.
  \item~Field and cluster galaxies are consistent in {\n} index, regardless of the cluster-centric radius.
 \end{itemize}  

For star-forming galaxies:
\begin{itemize}
  \item~We find that the mass normalised ({\M}$=10.7)$ average size of cluster star-forming galaxies within 0.5~{\Rv} is $16\pm{7}\%$ smaller, than field star-forming galaxies. However, this difference disappears if cluster star-forming galaxies are at a larger radius of 2~{\Rv}.
  \item~The fraction of galaxies with {\n} indices with $n\textgreater1$ for cluster star-forming galaxies within 0.5~{\Rv} is $50\%$ compared to a fraction of $30$ for field star-forming galaxies. Again, this difference disappears for a cluster sample at larger radius.
  \item~The mass-normalised average SFR of field star-forming galaxies is elevated by a factor of 2 (significance of $7-9\sigma$) compared to cluster star-forming galaxies, regardless of clustocentric radius. However, this trend appears to be driven by the high mass end, indicating that environment is more efficient at quenching galaxies with {\M}$\textgreater10.3$.
 \end{itemize}  

Our results are consistent with previous works which all show that the dependence of the {\ms} relation on environment for quiescent galaxies is minimal at best.
This continues to be surprising because quiescent galaxies are thought to be built up via mergers which should occur more frequently in clusters. 
These mergers are likely responsible, at least in part, for producing the morphology-density relation. 
It is, however, clear that the cluster environment plays an important role controlling the gas content of galaxies. 
Thus, perhaps it is more useful to study star-forming galaxies where environment can have its greatest effect on gas content, SFRs and sizes and possibly transition active galaxies into passive ones.

While there are few studies that examine the {\ms} relation of star-forming galaxies as a function of redshift, we can use those results to infer that environment does appear to influence the size of star-forming galaxies.
The lower average SFRs of cluster star-forming galaxies could mean that they have lost access to cold gas reservoirs in the cluster core and cannot grow via star-formation at the same rate as galaxies in the field.
Additional studies are needed to constrain the effects of environment on the growth mechanisms of star-forming galaxies, preferentially spanning $0\leq{z}\leq3$ so that it is clear at what epoch massive cluster star-forming galaxies are becoming quenched. 

\acknowledgments 
We thank the referee for their comprehensive and constructive comments. 
Research support to R.J.A is provided by the Australian Astronomical Observatory. 
G.G.K acknowledges the support of the Australian Research Council through the award of a Future Fellowship (FT140100933). 
K.G. acknowledges funding from the Australian Research Council (ARC) Discovery Program (DP) grant DP1094370.

\end{document}